\documentclass[final]{IEEEtran}
\usepackage{amsthm,amssymb,graphicx,multirow,amsmath,color,amsfonts,physics}%,ulem}
\usepackage[update,prepend]{epstopdf}
\usepackage[noadjust]{cite}
\usepackage{tikz}
\usepackage{bbm} % for \nbb1 
\usepackage{pdfpages}
\usepackage{balance}
\usepackage{multirow}
\usepackage{comment}
\usepackage{subfigure}
\allowdisplaybreaks % Allows breaking of eqnarray over multiple pages (avoids unnecessary blanks in the document before eqnarray)

\begin{document}
% Bold lowercase: syntax \nb# where # is {a ... z, 0,1}
\def\nba{{\mathbf{a}}}
\def\nbb{{\mathbf{b}}}
\def\nbc{{\mathbf{c}}}
\def\nbd{{\mathbf{d}}}
\def\nbe{{\mathbf{e}}}
\def\nbf{{\mathbf{f}}}
\def\nbg{{\mathbf{g}}}
\def\nbh{{\mathbf{h}}}
\def\nbi{{\mathbf{i}}}
\def\nbj{{\mathbf{j}}}
\def\nbk{{\mathbf{k}}}
\def\nbl{{\mathbf{l}}}
\def\nbm{{\mathbf{m}}}
\def\nbn{{\mathbf{n}}}
\def\nbo{{\mathbf{o}}}
\def\nbp{{\mathbf{p}}}
\def\nbq{{\mathbf{q}}}
\def\nbr{{\mathbf{r}}}
\def\nbs{{\mathbf{s}}}
\def\nbt{{\mathbf{t}}}
\def\nbu{{\mathbf{u}}}
\def\nbv{{\mathbf{v}}}
\def\nbw{{\mathbf{w}}}
\def\nbx{{\mathbf{x}}}
\def\nby{{\mathbf{y}}}
\def\nbz{{\mathbf{z}}}
\def\nb0{{\mathbf{0}}}
\def\nb1{{\mathbf{1}}}

% Bold capital letters: syntax \nb# where # is {A ... Z}
\def\nbA{{\mathbf{A}}}
\def\nbB{{\mathbf{B}}}
\def\nbC{{\mathbf{C}}}
\def\nbD{{\mathbf{D}}}
\def\nbE{{\mathbf{E}}}
\def\nbF{{\mathbf{F}}}
\def\nbG{{\mathbf{G}}}
\def\nbH{{\mathbf{H}}}
\def\nbI{{\mathbf{I}}}
\def\nbJ{{\mathbf{J}}}
\def\nbK{{\mathbf{K}}}
\def\nbL{{\mathbf{L}}}
\def\nbM{{\mathbf{M}}}
\def\nbN{{\mathbf{N}}}
\def\nbO{{\mathbf{O}}}
\def\nbP{{\mathbf{P}}}
\def\nbQ{{\mathbf{Q}}}
\def\nbR{{\mathbf{R}}}
\def\nbS{{\mathbf{S}}}
\def\nbT{{\mathbf{T}}}
\def\nbU{{\mathbf{U}}}
\def\nbV{{\mathbf{V}}}
\def\nbW{{\mathbf{W}}}
\def\nbX{{\mathbf{X}}}
\def\nbY{{\mathbf{Y}}}
\def\nbZ{{\mathbf{Z}}}

% \mathcal: syntax \ncal# where # is {A ... Z}
\def\ncalA{{\mathcal{A}}}
\def\ncalB{{\mathcal{B}}}
\def\ncalC{{\mathcal{C}}}
\def\ncalD{{\mathcal{D}}}
\def\ncalE{{\mathcal{E}}}
\def\ncalF{{\mathcal{F}}}
\def\ncalG{{\mathcal{G}}}
\def\ncalH{{\mathcal{H}}}
\def\ncalI{{\mathcal{I}}}
\def\ncalJ{{\mathcal{J}}}
\def\ncalK{{\mathcal{K}}}
\def\ncalL{{\mathcal{L}}}
\def\ncalM{{\mathcal{M}}}
\def\ncalN{{\mathcal{N}}}
\def\ncalO{{\mathcal{O}}}
\def\ncalP{{\mathcal{P}}}
\def\ncalQ{{\mathcal{Q}}}
\def\ncalR{{\mathcal{R}}}
\def\ncalS{{\mathcal{S}}}
\def\ncalT{{\mathcal{T}}}
\def\ncalU{{\mathcal{U}}}
\def\ncalV{{\mathcal{V}}}
\def\ncalW{{\mathcal{W}}}
\def\ncalX{{\mathcal{X}}}
\def\ncalY{{\mathcal{Y}}}
\def\ncalZ{{\mathcal{Z}}}

% \mathbb: syntax \nbb# where # is {A ... Z}
\def\nbbA{{\mathbb{A}}}
\def\nbbB{{\mathbb{B}}}
\def\nbbC{{\mathbb{C}}}
\def\nbbD{{\mathbb{D}}}
\def\nbbE{{\mathbb{E}}}
\def\nbbF{{\mathbb{F}}}
\def\nbbG{{\mathbb{G}}}
\def\nbbH{{\mathbb{H}}}
\def\nbbI{{\mathbb{I}}}
\def\nbbJ{{\mathbb{J}}}
\def\nbbK{{\mathbb{K}}}
\def\nbbL{{\mathbb{L}}}
\def\nbbM{{\mathbb{M}}}
\def\nbbN{{\mathbb{N}}}
\def\nbbO{{\mathbb{O}}}
\def\nbbP{{\mathbb{P}}}
\def\nbbQ{{\mathbb{Q}}}
\def\nbbR{{\mathbb{R}}}
\def\nbbS{{\mathbb{S}}}
\def\nbbT{{\mathbb{T}}}
\def\nbbU{{\mathbb{U}}}
\def\nbbV{{\mathbb{V}}}
\def\nbbW{{\mathbb{W}}}
\def\nbbX{{\mathbb{X}}}
\def\nbbY{{\mathbb{Y}}}
\def\nbbZ{{\mathbb{Z}}}

% \mathfrak:
\def\nfrakR{{\mathfrak{R}}}

% Roman: {\rm } syntax \nrm# where # is {a ... z}
\def\nrma{{\rm a}}
\def\nrmb{{\rm b}}
\def\nrmc{{\rm c}}
\def\nrmd{{\rm d}}
\def\nrme{{\rm e}}
\def\nrmf{{\rm f}}
\def\nrmg{{\rm g}}
\def\nrmh{{\rm h}}
\def\nrmi{{\rm i}}
\def\nrmj{{\rm j}}
\def\nrmk{{\rm k}}
\def\nrml{{\rm l}}
\def\nrmm{{\rm m}}
\def\nrmn{{\rm n}}
\def\nrmo{{\rm o}}
\def\nrmp{{\rm p}}
\def\nrmq{{\rm q}}
\def\nrmr{{\rm r}}
\def\nrms{{\rm s}}
\def\nrmt{{\rm t}}
\def\nrmu{{\rm u}}
\def\nrmv{{\rm v}}
\def\nrmw{{\rm w}}
\def\nrmx{{\rm x}}
\def\nrmy{{\rm y}}
\def\nrmz{{\rm z}}

% Special symbols
\def\nbydef{:=}
\def\nborel{\ncalB(\nbbR)}
\def\nboreld{\ncalB(\nbbR^d)}
\def\sinc{{\rm sinc}}

% Theorems etc.
\newtheorem{lemma}{Lemma}
\newtheorem{thm}{Theorem}
\newtheorem{definition}{Definition}
\newtheorem{ndef}{Definition}
\newtheorem{nrem}{Remark}
\newtheorem{theorem}{Theorem}
\newtheorem{prop}{Proposition}
\newtheorem{cor}{Corollary}
\newtheorem{example}{Example}
\newtheorem{remark}{Remark}
\newtheorem{assumption}{Assumption}
	
%%%%%%%% Backwards compatibility

\newcommand{\ceil}[1]{\lceil #1\rceil}
\def\argmin{\operatorname{arg~min}}
\def\argmax{\operatorname{arg~max}}
\def\figref#1{Fig.\,\ref{#1}}%
\def\E{\mathbb{E}}
\def\EE{\mathbb{E}^{!o}}
\def\P{\mathbb{P}}
\def\pc{\mathtt{P_c}}
\def\rc{\mathtt{R_c}}   % rate coverage
\def\p{p}

\def\V{\operatorname{Var}}
\def\erfc{\operatorname{erfc}}
\def\erf{\operatorname{erf}}
\def\opt{\mathrm{opt}}
\def\R{\mathbb{R}}
\def\Z{\mathbb{Z}}

\def\LL{\mathcal{L}^{!o}}
\def\var{\operatorname{var}}
\def\supp{\operatorname{supp}}

\def\N{\sigma^2}
\def\T{\beta}							% Threshold = \beta_i
\def\sinr{\mathtt{SINR}}			% Signal to interference plus noise ratio
\def\snr{\mathtt{SNR}}
\def\sir{\mathtt{SIR}}
\def\ase{\mathtt{ASE}}
\def\se{\mathtt{SE}}

\def\calN{\mathcal{N}}
\def\FE{\mathcal{F}}
\def\calA{\mathcal{A}}
\def\calK{\mathcal{K}}
\def\calT{\mathcal{T}}
\def\calB{\mathcal{B}}
\def\calE{\mathcal{E}}
\def\calP{\mathcal{P}}
\def\calL{\mathcal{L}}
%\DeclareMathOperator{\Tr}{Tr}
%\DeclareMathOperator{\rank}{rank}
%\DeclareMathOperator{\Pois}{Pois}

%\DeclareMathOperator{\TC}{\mathtt{TC}}
%\DeclareMathOperator{\TCL}{\mathtt{TC_l}}
%\DeclareMathOperator{\TCU}{\mathtt{TC_u}}

% Fading
\def\l{\ell}
\newcommand{\fad}[2]{\ensuremath{\mathtt{h}_{#1}[#2]}}
\newcommand{\h}[1]{\ensuremath{\mathtt{h}_{#1}}}

\newcommand{\err}[1]{\ensuremath{\operatorname{Err}(\eta,#1)}}
\newcommand{\FD}[1]{\ensuremath{|\mathcal{F}_{#1}|}}

%% Symbols changed
% \def\i{\mathbf{1}}					% changed to \nb1
% \def\d{\mathrm{d}}					% changed to \nrmd
% \def\L{\mathcal{L}}					% changed to \ncalL
% \begin{definition}					% changed to \begin{ndef}

% \l also gives problems. Use \ell after defining it if needed.

%% D2D def
\def\Bx{{\mathcal{B}}^x}
\def\Bxx{{\mathcal{B}}^{x_0}}
\def\jx{y}
\def\m{(\bar{n}-1)}
\def\mm{\bar{n}-1}
\def\Nx{{\mathcal{N}}^x}
\def\Nxo{{\mathcal{N}}^{x_0}}
\def\wj{w_{jx_0}}
\def\uij{u_{jx}}
 \def\yj{y}
 \def\yjx{y}
 \def\zjx{z_x}
 \def \tx {y_0}
 \def \htx {h_0}

\def\rx{z_{1}}
\def\ry{z_{2}}

\def\Rx{Z_{1}}
\def\Ry{Z_{2}}

%% fading
\def \hyxx {h_{y_{x_0}}}
\def \hyx {h_{y_x}}

\def\nbb1{\mathbbm{1}}
\def\xi{x_i}
\def\xj{x_j}
\def\xx{x_0}
\def\yk{y_k}
\def\yy{y_0}
\def\ie{{\em i.e. }}
\def\eg{{\em e.g. }}
\def\iid{{\em i.i.d. }}
\def\avg{\rm avg}

\def\rmnuma{\rm\uppercase\expandafter{\romannumeral1}}
\def\rmnumb{\rm\uppercase\expandafter{\romannumeral2}}
\def\rmnumc{\rm\uppercase\expandafter{\romannumeral3}}
\def\rmnumd{\rm\uppercase\expandafter{\romannumeral4}}
\def\rmnume{\rm\uppercase\expandafter{\romannumeral5}}
\def\rmnumf{\rm\uppercase\expandafter{\romannumeral6}}
\pagenumbering{gobble}
\graphicspath{{./Figures/}}
\title{
Energy-as-a-Service for RF-Powered IoE Networks: A Percolation Theory Approach}
%Large-Scale IoE Connectivity  of RF-powered IoE: A Percolation Theory Approach}
\author{
 Hao Lin, Ainur Zhaikhan, Mustafa A. Kishk, Hesham ElSawy, and Mohamed-Slim Alouini
\thanks{Hao Lin is with the Electrical and Computer Engineering Program, Computer, Electrical and Mathematical Sciences and Engineering Division (CEMSE), King Abdullah University of Science and Technology (KAUST), Thuwal 23955-6900, Kingdom of Saudi Arabia (e-mail: hao.lin.std@gmail.com).\\
\indent A. Zhaikhan is
with {\'E}cole Polytechnique F{\'e}d{\'e}rale de Lausanne (EPFL), Lausanne 1015,
Switzerland. (e-mail: ainur.zhaikhan@epfl.ch).\\
\indent Mustafa A. Kishk is with the Department of Electronic Engineering,
Maynooth University, Maynooth, W23 F2H6 Ireland (e-mail:
mustafa.kishk@mu.ie).\\
\indent Hesham ElSawy is with the School of Computing, Queen’s University, ON, Canada K7L 3N6 (e-mail: hesham.elsawy@queensu.ca).\\
\indent Mohamed-Slim Alouini is with the CEMSE Division, King Abdullah
University of Science and Technology (KAUST), Thuwal 23955-6900,
Saudi Arabia (e-mail: slim.alouini@kaust.edu.sa).}
}

\maketitle
% \vspace{-2cm}
%\thispagestyle{empty}
%\pagestyle{empty}
\begin{abstract}
Due to the involved massive number of devices, radio frequency (RF) energy harvesting is indispensable to realize the foreseen Internet-of-Everything (IoE) within 6G networks. Analogous to the cellular networks concept, shared energy stations (ESs) are foreseen to supply energy-as-a-service (EaaS) in order to recharge devices that belong to different IoE operators who are offering diverse use cases. Considering the capital expenditure (CAPEX) for ES deployment along with their finite wireless energy transfer (WET) zones, spatial energy gaps are plausible. Furthermore, the ESs deployment cannot cover $100\%$ of the energy-harvesting devices of all coexisting IoE use cases. In this context, we utilize percolation theory to characterize the feasibility of large-scale device-to-device (D2D) connectivity of IoE networks operating under EaaS platforms. Assuming that ESs and IoE devices follow independent Poisson point processes (PPPs), we construct a connectivity graph for the IoE devices that are within the {WET zones} of ESs. Continuum percolation on the construct graph is utilized to derive necessary and sufficient conditions for large-scale RF-powered D2D connectivity in terms of the required IoE device density and communication range along with the required ESs density and WET zone size. Fixing the IoE network parameters along with the {size of WET zones}, we obtain the approximate critical value of the ES density that ensures large-scale connectivity using the inner-city and Gilbert disk models. By imitating the bounds and combining the approximations, we construct an approximate expression for the critical ES density function, which is necessary to minimize the EaaS CAPEX under the IoE connectivity constraint. 

\end{abstract}
% \vspace{-0.3cm}
\begin{IEEEkeywords}
% \vspace{-0.3cm}
Energy-as-a-Service (EaaS), Internet of Everything (IoE),  stochastic geometry, percolation theory, Gilbert disk model.%energy harvesting,,graph theory,
\end{IEEEkeywords}

\section{Introduction} \label{sec:Intro}

In the 6G era, wireless connectivity is foreseen to be global, ubiquitous, and massive. In addition to transforming every object in our life into an intelligent and connected device, the proliferating 6G Internet of Everything (IoE) use cases (e.g., digital twins, ubiquitous environmental monitoring, precision agriculture, smart cities, wearable devices, etc.) will bring massive numbers of additional wireless devices~\cite{9509294,9627726,9802635}. In such a massive network setup, monitoring and replacing/recharging the batteries of the devices is an overwhelming process that imposes high operational expenditure (OPEX). Such high OPEX can be alleviated by utilizing wireless energy transfer (WET) along with energy-harvesting (EH) technologies~\cite{WET1, bi2016wireless,6951347,7120022}. In particular, energy providers deploy energy stations (ESs) for WET, which offers energy-as-a-service (EaaS) to IoE operators/consumers who equip their devices with EH modules. The range of far-field wireless charging can reach tens of meters to meet the charging requirements of most mobile application scenarios. Also, the use of green energy can flexibly provide free and continuous electromagnetic energy for ESs \cite{9627726,9802635}. The proliferating IoE use cases make such energy services a worthwhile business opportunity \cite{paukstadt2021energy}. In fact, the EaaS platform is considered an attractive business model that is estimated to exceed $\$220$ billion annual global market value by 2026 \cite{EaaS}. In the future, such EaaS platforms can not only serve mobile users and sensor networks on the ground, but also provide wireless power support for unmanned aerial vehicles (UAVs) and more near space platforms \cite{10580989}.\\
% I wish you the best of success.
\indent To reduce CAPEX, EaaS operators strive to satisfy the IoE energy demand with minimal ESs deployment. Assuming that IoE operators/consumers mandate large-scale D2D connectivity of their devices,\footnote{D2D connectivity can reflect the maximum communication range of each device, and hence, large-scale connectivity enables long-range information dissemination. D2D connectivity can also be interpreted as the maximum sensing range of each device, and hence, large-scale connectivity is important for monitoring and intrusion detection IoE use cases.} graph and percolation theories are best suited to estimate the necessary and sufficient ES deployments that satisfy the IoE operators/consumers connectivity demand. In particular, graph theory provides a tractable mathematical representation for the interactions among network components, where vertices are used to model devices and edges represent direct relationships among vertices \cite{west2001introduction,Diestel_2017}. Percolation theory is used to characterize the large-scale connectivity of graphs. Since D2D connectivity and WET zone coverage depend on the relative locations of IoE devices with respect to each other and with respect to ESs, random geometric graphs (RGGs) are utilized~\cite{haenggi2012stochastic}. Assuming that ESs and IoE devices follow independent Poisson point processes (PPPs), a WET-aware connectivity (WC)-RGG for the IoE devices is constructed based on their relative locations, D2D communication range, and WET zone coverage. Using continuum percolation on the IoE WC-RGG, we characterize the network parameters that enable large-scale D2D connectivity of the IoE network. Hence, helping operators estimate the required CAPEX to deliver a successful EaaS. 

\subsection{Related Work}
In this paper, we propose to apply the EaaS platform to support large-scale D2D communications between RF-powered IoE devices. Using percolation theory, we prove the feasibility of such an EaaS platform and help operators estimate critical ES density, further, the required CAPEX. Therefore, we divide the related work into: i) percolation theory for wireless networks and ii) stochastic geometry for RF-powered wireless networks.

\indent \textit{Percolation theory for wireless networks:} Continuum percolation on RGGs has been widely utilized to study the large-scale network-wide performance of wireless networks with randomly located devices~\cite{haenggi2012stochastic, 5226957,elsawy2023tutorial}. For instance, the authors in \cite{SINR1} utilized continuum percolation to quantify the impact of aggregate network interference on large-scale network connectivity. End-to-end capacity in large-scale networks was characterized in \cite{SINR2} via percolation theory. Using the concept of information secrecy, the authors in \cite{Pinto1, Goel} found the relative densities of legitimate devices and eavesdroppers that enable large-scale private message dissemination among legitimate devices. In cognitive radio networks, the authors in \cite{Cog1, Cog2, Cog3} characterized the relationship between the intensity of secondary devices and primary devices to allow large-scale connectivity of the secondary network. Simultaneous large-scale connectivity of both the primary and secondary devices in cognitive networks was studied in \cite{yemini2019simultaneous}. For sensing/monitoring applications, the path exposure problem was characterized in \cite{8794718, sensing}, where the authors found the critical density of sensors/cameras to detect moving objects over arbitrary paths through a given region. In the context of cyber-security, the authors in \cite{9240972} utilized continuum percolation to quantify the required spatial firewalls to thwart large-scale outbreaks of malware worms in wireless networks. By investigating the BS networks assisted by reconfigurable intelligent surfaces (RISs), authors in \cite{wu2023connectivity} derived the lower bound of the critical density of RISs. In \cite{han2024dynamic}, based on dynamic bond percolation, authors proposed an evolution model to characterize the reliable topology evolution affected by the nodes and links states. However, WET-enabled large-scale connectivity of RF-powered IoE networks is still an open problem.\\
\indent \textit{Stochastic geometry for
RF-powered wireless networks:} RF-powered wireless communication has been extensively studied at the link-level performance as summarized in~\cite{7010878}. In large-scale RF-powered networks, stochastic geometry was utilized to provide spatial averages of link-level performance metrics such as coverage probability, rate, or delay. For instance, the work in \cite{6786061, 9134826} characterized the energy efficiency for PPP downlink cellular networks that are powered via renewable energy sources. Outage probability and rate of downlink simultaneous wireless information and power transfer (SWIPT) were studied in \cite{9723559} for cooperative non-orthogonal multiple access (NOMA) and in \cite{9201540} for cell-free massive multiple-input multiple-output (MIMO) networks. On the uplink side, the work in \cite{Sakr_EH} characterized coverage probability with channel inversion power control. Delay and packet throughput for grant-free uplink access of RF-powered Internet of Things (IoT) was characterized in \cite{Gharbieh_EH}. Outage probability and self-sustainability of D2D RF-powered IoT networks were characterized in \cite{Fatma_EH, 9093022, 9463400}. End-to-end outage probability for RF-powered D2D decode-and-forward relaying was studied in \cite{8888216}. D2D-enabled proactive caching for RF-powered wireless networks was studied in \cite{8896898}. For non-terrestrial networks, the authors in \cite{9390298} optimized the altitude of RF-powered aerial base stations serving terrestrial devices. A similar scenario was studied in \cite{Kishk1} for laser-powered aerial base stations. Authors in \cite{9507551} combined the backscatter communication with RF-powered cognitive networks, and investigated the resource allocation for channel selection and backscatter power allocation. In \cite{10036459}, authors evaluated the effect on RF-powered IoT network performance of RIS density, RIS reflecting element number, and charging stage ratio. However, none of the aforementioned stochastic geometry frameworks can be utilized to study network-wide connectivity of RF-powered IoE networks in order to help EaaS providers estimate the required intensity of ESs.
\subsection{Contributions}
\indent As an important performance metric of IoE networks, connectivity has its unique value. It refers to the ability to generate remote multi-hop D2D transmission links, which further promotes the flexibility and reliability of the IoE networks. Higher network-wide connectivity of IoE networks not only helps realize more functions in different applications, but also helps achieve better performance, such as higher detection accuracy, stronger fault tolerance, and stronger noise immunity et al. However, to the best of the authors' knowledge, there are no available network-wide connectivity studies for RF-powered wireless networks, which is foundational to estimating the CAPEX for EaaS providers. This paper covers such a gap with the following set of contributions: 
\begin{itemize}
    \item We develop a novel WC-RGG, and utilize continuum percolation, to study large-scale multi-hop D2D connectivity in RF-powered IoE networks.
    \item We prove the existence of a finite critical density of ESs that enable network-wide connectivity of battery-less IoE devices.
    \item We obtain an accurate approximate expression for critical ES density as a function of IoE device density, IoE D2D communication range, and ES WET zone sizes, which is important to estimate the CAPEX of EaaS. 
    \item Our analysis provides an alternative way to compute the well-celebrated approximate percolation critical density for PPP networks with unit-length connectivity radius as $\lambda_c(1)=\frac{4\ln 2}{\pi}\approx 0.883$, which further validates our analysis.  
\end{itemize}

\textbf{Organization:} The organization of the paper is as follows. The system model and assumptions are detailed in Sec.II. Sec.III proves the feasibility of large-scale connectivity of RF-powered IoE networks under EaaS platforms. Sec.V provides useful approximations for the critical density of ESs, which give insights into the CAPEX of EaaS. Numerical and simulation results are discussed in Sec.V before the paper is concluded in Sec.VI.

\textbf{Notations:} In this paper, we utilize the following set of notations. The probability of an event is denoted as $\mathbb{P}\{\cdot\}$. 
The right arrow $\rightarrow$ means the event on the left of it causes events on the right side. The main symbols used in this paper are summarized in Table.\ref{tab:TableOfNotations}.

\begin{table*}[t]\caption{Table of Notations}
\centering
\begin{center}
% \resizebox{\textwidth}{!}{
\renewcommand{\arraystretch}{1.4}%1.4
    \begin{tabular}{ {c} | {l} }
    \hline
        \hline
    \textbf{Notation} & \textbf{Description} \\ \hline
     $\Phi$; $\lambda_r$; $r_r$ & The set of locations of IoE devices; the density of IoE devices; the maximum communication range \\ \hline
     $\Psi$; $\lambda_f$; $r_f$ & The set of locations of ESs; the density of ESs; the maximum charging/energy supporting range \\ \hline
     $G(V,E)=\{V,E\}$ & The RGG representation of active IoE devices with vertex set $V$ and edge set $E$ \\ \hline
     $\theta(\lambda_r,r_r,\lambda_f,r_f)$ & The probability of percolation in $G(V,E)$ when $\lambda_f<\infty$ \\ \hline
     $\theta(\lambda_r,r_r)$ & The probability of percolation in $G(V,E)$ when $\lambda_f=\infty$ \\ \hline
     $\ncalL_h^L$; $\ncalH_L$; $a_L$; $\ncalE_{out}^L$ & The hexagonal lattice, randomly selected hexagon, side length of a hexagon and outer envelope in sub-critical regime \\ \hline
     $\ncalL_h^U$; $\ncalH_U$; $a_U$; $\ncalE_{in}^U$ & The hexagonal lattice, randomly selected hexagon, side length of a hexagon and inner envelope in super-critical regime \\ \hline
     $K$; $K_{\ncalL_h^L}$; $K_{\ncalL_h^U}$ & The connected component in $G$, $\ncalL_h^L$ and $\ncalL_h^U$ \\ \hline
     $\lambda_f^c$; $\lambda_f^L$; $\lambda_f^U$ & The critical ES density and its lower bound and upper bound \\ \hline
     $\lambda_f^{IC}$; $\lambda_f^{GD}$; $\lambda_f^*$ & The approximations using the Inner-city model, a simple Gilbert model or combining them together \\ \hline
     \hline
    \end{tabular}
    % }
\end{center}
\label{tab:TableOfNotations}
%\vspace{-8mm}
\end{table*}

\section{System Model} \label{sec:SysMod}
\begin{figure}
    \centering
    \includegraphics[width=0.95\columnwidth]{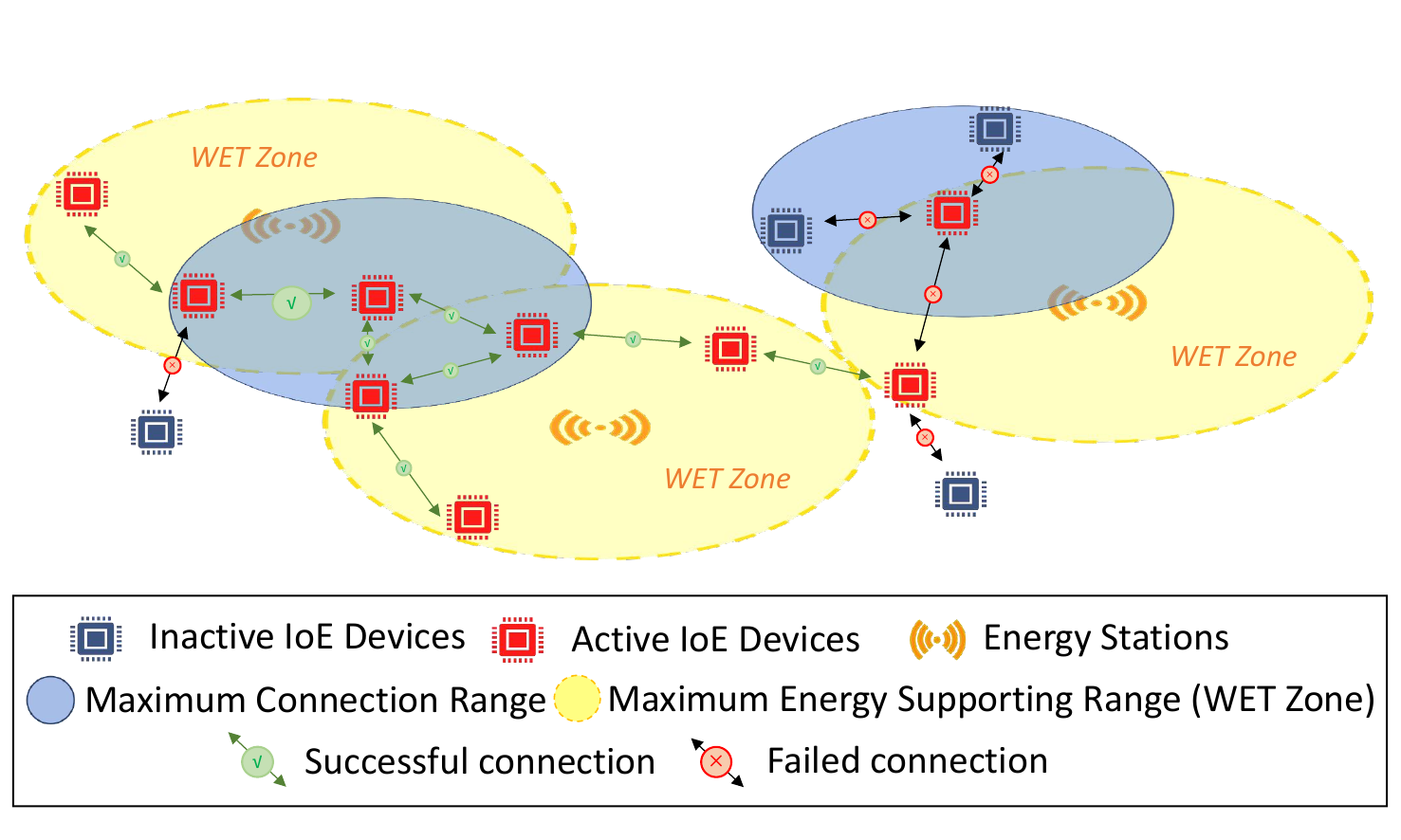}
    \caption{Illustration for RF-powered IoE operation under the EaaS platform. The IoE devices inside the WET zones of ESs can be activated (marked in red), while other IoE devices are inactive (marked in blue). A successful connection between two active IoE devices can be achieved when the distance between them is smaller than the maximum connection range.}
    \label{fig:energycharging}
\end{figure}

We consider an IoE network with battery-less devices that are spatially scattered according to a PPP $\Phi=\{x_0, x_1, \cdots, x_i, \cdots\}\subset \R^2$ with density $\lambda_r$. The IoE devices are equipped with EH modules that are powered through an EaaS operator with ESs that are deployed according to an independent PPP $\Psi=\{y_0,y_1,\cdots,y_k,\cdots\}\subset \R^2$ with density $\lambda_f$. The independence between $\Phi$ and $\Psi$ is justified by the fact that an EaaS operator is serving many IoE networks, and hence, the ES deployment is not customized to any of them. Instead, the EaaS operators deploy their ESs in strategic locations where there is high demand from several IoE network operators/consumers. The ESs have a maximum WET range $r_f$ (i.e., the maximum charging range for IoE devices), where each ES can provide EaaS to activate all IoE devices inside their WET zones. The IoE operators mandate a long-range network-wide D2D connectivity among their battery-less devices. In particular, the IoE operators seek an EaaS platform that is able to activate a sufficient number of devices that span the spatial domain and can reach each other through multi-hop D2D connections. Two battery-less IoE devices can establish a D2D connectivity if and only if they satisfy the following two conditions; i) each of the IoE devices falls within a WET zone of at least one ESs, and ii) both IoE devices are within the D2D communication range of $r_r$ from each other. A pictorial illustration of the RF-powered IoE operation under the EaaS platform is shown in Fig.\ref{fig:energycharging}. Given the stringent power and complexity constraints of IoE devices, it is plausible to assume that $r_r\leq r_f$, where each ES can support enough IoE devices and form a small IoE network around it. The transmit power of each ES should be much higher than the transmit power of each IoE device because the maximum distance between any ES and IoE devices in its WET zone should be larger than the distance between adjacent IoEs, and each ES is responsible for the energy supply of multiple IoE devices.\\
\indent Following the aforementioned two connectivity conditions for battery-less devices, we construct a WC-RGG for RF-powered IoE networks operating under an EaaS platform. The WC-RGG is denoted as $G=\{V,E\}$, where $V$ is the set of vertices defined as:
\begin{equation}
    V=\{\xi\in \Phi:\min_{\yk\in\Psi}\|\xi-\yk\|\leq r_f\},
\end{equation}
and $E$ is the set of edges given by:
\begin{equation}
    E=\{\overline{\xi\xj}:\|\xi-\xj\|\leq r_r,\ \xi,\xj\in V\}.
\end{equation}

\indent Large-scale network-wide connectivity of the RF-powered IoE can be characterized via continuum percolation over the WC-RGG.\footnote{It is important to mention that continuum percolation does not guarantee full connectivity of the WC-RGG. Instead, continuum percolation guarantees that there exists a large number of vertices within the WC-RGG that span the spatial domain and can reach each other through multi-hop connections.} We first define a component $K\subseteq G$, as a connected sub-graph where any two vertices within $K$ are connected through a set of consecutive edges in $E$. The non-zero percolation is an event where there exists an infinite-size connected component $K$ within the WC-RGG with a non-zero probability. Without loss of generality, we assume that there is a device located at the origin that belongs to the connected component $K(0)\subseteq G(V,E)$. The set cardinality of $K(0)$ is denoted as $|K(0)|$ and the percolation probability in $G$ is defined as:
\begin{equation}
    \theta(\lambda_r,r_r,\lambda_f,r_f)=\P\{|K(0)|=\infty\}.
\end{equation}
\indent To minimize the CAPEX of the EaaS operator and satisfy the large-scale D2D connectivity mandate of the IoE operator/consumer, the design objective for the ESs deployment can be formulated as:
\begin{equation}
    \begin{array}{ll}
       \text{ minimize}  & \lambda_f  \\
       \text{ subject to}  & \theta(\lambda_r,r_r,\lambda_f,r_f)>0.  \\
    \end{array}
    \label{perpro}
\end{equation}
\indent The problem formulation in \eqref{perpro} seeks the minimum density of ESs that ensures large-scale D2D connectivity of the battery-less IoE networks. In the extreme case, when $\lambda_f$ approaches $\infty$, the ESs are dense enough to activate all IoE devices. Such a highly dense ES regime captures a scenario where there is no energy scarcity and is denoted hereafter by an infinity subscript to highlight that $\lambda_f$ approaches $\infty$. In the highly dense ES regime, the WC-RGG transforms to a conventional RGG with vertices $V_\infty=\Phi$ and edges
\begin{equation}
    E_\infty=\{\overline{\xi\xj}:\|\xi-\xj\|\leq r_r,\ \xi,\xj\in \Phi\}.
\end{equation}
\indent In the highly dense ES regime, the percolation probability is only relevant to the IoE parameters $\lambda_r$ and $r_r$, which is given by:
\begin{equation}
    \theta_\infty(\lambda_r,r_r)=\P\{|K(0)|=\infty\}.
\label{mperpro}
\end{equation}
\indent Since the highly dense ESs regime transforms the WC-RGG to a conventional RGG of a PPP with density $\lambda_r$ a connectivity range $r_r$, the critical IoE parameters that ensure continuum percolation in such regime can be directly obtained from \cite{meester1996continuum} as   
\begin{equation}
    \lambda_r>\frac{\lambda_c(1)}{r_r^2},
\label{basiccondition}
\end{equation}
where $0.768<\lambda_c(1)<3.37$ is the percolation critical density for a PPP RGG with unit-length connectivity radius.\footnote{It is worth noting that the exact value for $\lambda_c(1)$ is still an open problem. There are only known bounds and approximations for the critical density $\lambda_c(1)$.} It is worth noting that \eqref{basiccondition} is not only the sufficient condition for non-zero percolation when $\lambda_f$ approaches $\infty$, but also the necessary condition in RF-powered IoE networks where $\lambda_f<\infty$.

\section{Proof of concept}
\label{sec:concept}
In this section, we define and prove the phase transition of the percolation probability for the WC-RGG (i.e., $G(V,E)$). In particular, when the density of IoE devices satisfies the condition (\ref{basiccondition}), there exists a critical ES density $\lambda_f^c$, below which remote multi-hop links can not be established and $G(V,E)$ operates in the sub-critical regime. On the contrary, when the density of the ESs is above $\lambda_f^c$, then $G(V,E)$ operates in the super-critical regime, where remote multi-hop links can be realized from the origin with a non-zero probability.\\
\indent Following the percolation theory convention, we organize the proof into the following subsections. Sec.\ref{subsec:sub} presents the sub-critical regime for $G(V,E)$, which proves the need for enough ES density. We prove the sufficient condition for no percolation and derive the lower bound of $\lambda_f^c$. Sec.\ref{subsec:sup} presents the super-critical regime, in which we prove the sufficient condition for non-zero percolation and derive the upper bound of $\lambda_f^c$. Last but not least, Sec.\ref{subsec:cri} completes the proof of phase transition by showing the monotonicity of percolation probability in ES density $\lambda_f$. In Sec.\ref{subsec:cri}, we also derive the relationship between critical ES density and IoE device density, and define the critical ES density function. 
\subsection{Sub-critical regime}
\label{subsec:sub}
\begin{figure}[htbp]
\centering
\subfigure[Inactive faces.]{
\begin{minipage}[t]{0.9\linewidth}
\centering 
\includegraphics[width=0.8\textwidth]{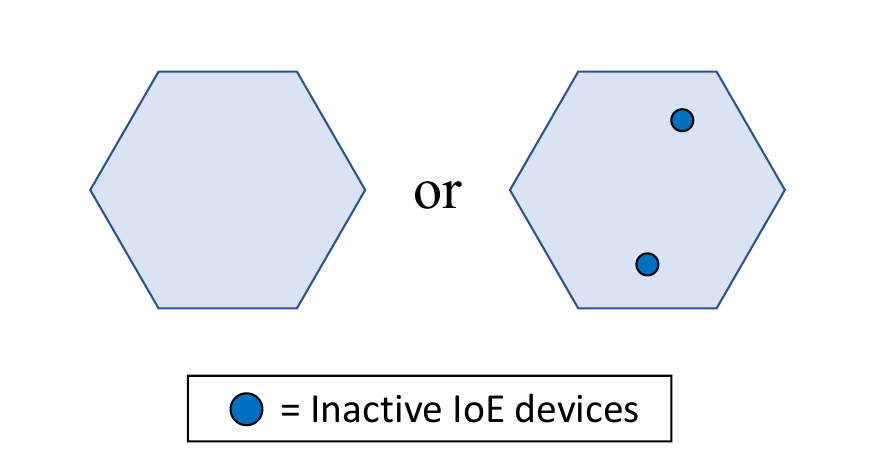}
\label{fig:ClosedFace}
\end{minipage}}

\subfigure[Outer envelope.]{
\begin{minipage}[t]{0.9\linewidth}
\centering 
\includegraphics[width=1\textwidth]{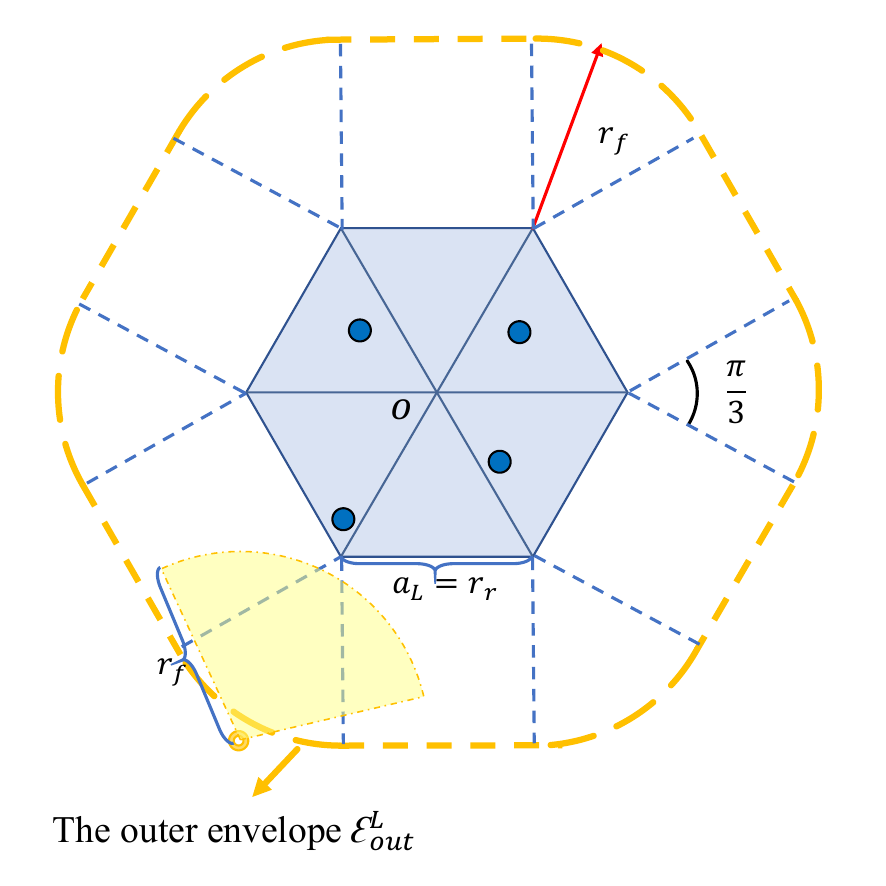}
\label{fig:outerenvelope}
\end{minipage}}

\subfigure[An inactive circuit.]{
\begin{minipage}[t]{0.9\linewidth}
\centering 
\includegraphics[width=1\textwidth]{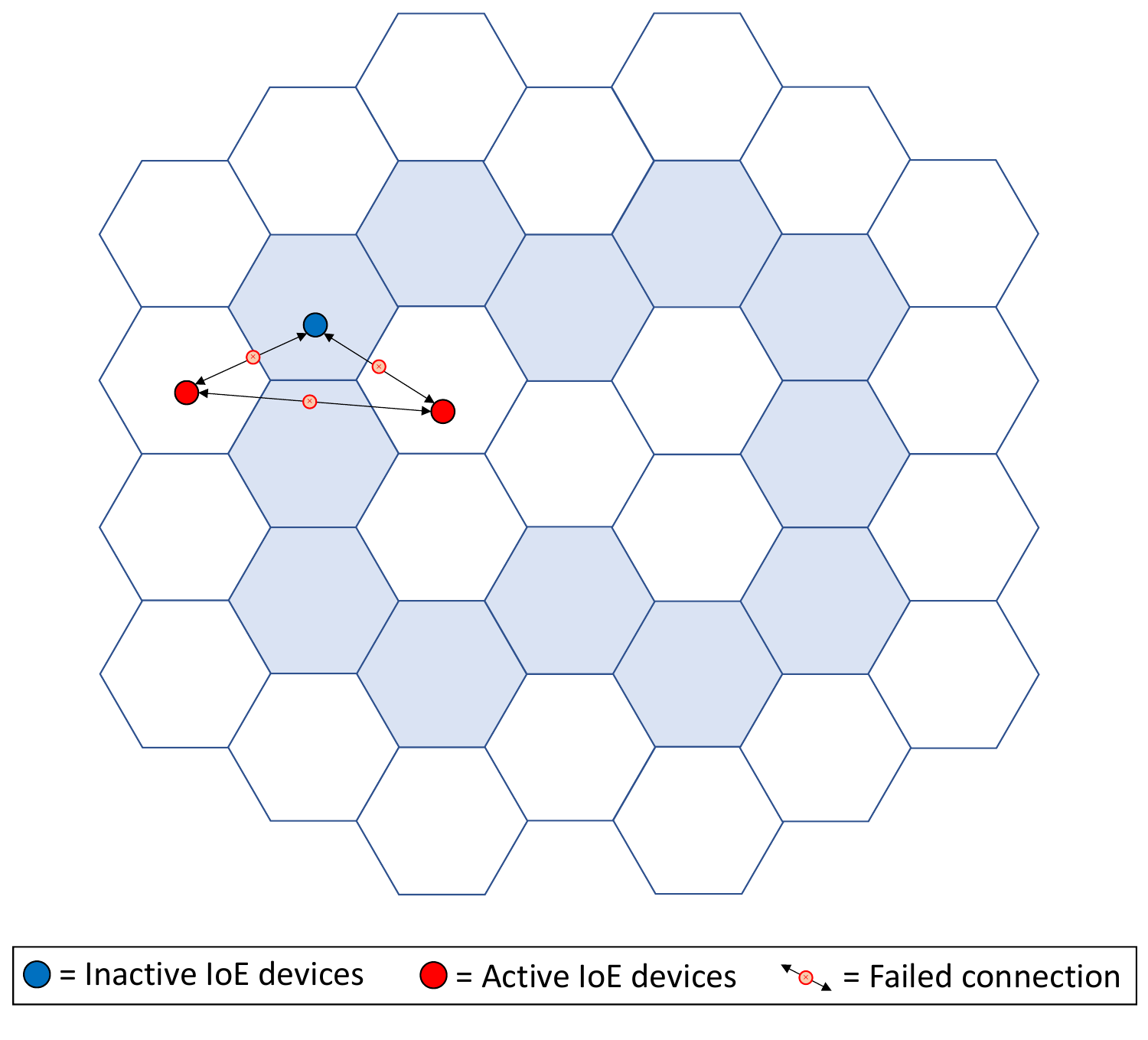}
\label{fig:ClosedCircuit}
\end{minipage}}
\caption{Illustration for the inactive faces, outer envelope, and inactive circuit. As shown in Fig.\ref{fig:ClosedFace}, a hexagon without any active IoE devices is considered an inactive face. Fig.\ref{fig:outerenvelope} is an extreme case we use to derive the lower bound. If there is no ES inside the outer envelope, all IoE devices in the hexagon are inactive. Fig.\ref{fig:ClosedCircuit} shows that an inactive circuit appears when $\P\{\ncalH_L \rm{\ is\ inactive}\}>\frac{1}{2}$. Because $r_r=a_L$, the IoE device inside the circuit can not be connected with those outside the circuit.}
\end{figure}

This section proves that large-scale network-wide multi-hop links can not be realized when the ESs are not dense enough. In this case, the EaaS is not successful. We find the sufficient condition for no percolation and derive the lower bound of critical ES density $\lambda_f^c$. As a common practice in percolation theory, we study continuum percolation by mapping it to discrete lattices. Inspired by \cite{9240972} and \cite{pinto2012percolation}, we analyze the sub-critical regime by mapping the WC-RGG $G(V,E)$ to a hexagonal lattice. We consider the special scenario where $r_r$ is the same as the side length of the hexagon such that two active IoE devices in two nonadjacent hexagons can not be connected via a direct D2D link.  \\
\textbf{Mapping to a Hexagonal Lattice in Sub-critical Regime:} Let $\ncalL_h^L$ be a hexagonal lattice with a side $a_L=r_r$ and let $\ncalH_L$ be a randomly selected hexagon in the lattice. Hereafter, $\ncalH_L$ is denoted as a face in $\ncalL_h^L$. Such a setup implies that $a_L$ is the minimum distance between IoE devices in two nonadjacent hexagons. 
\begin{definition}[Inactive face in sub-critical regime] A face $\ncalH_L$ is denoted as inactive if it contains no edges of $E$. As shown in Fig.\ref{fig:ClosedFace}, the conditions to ensure that $\ncalH_L$ is an inactive face are: i) there is no IoE device in $\ncalH_L$ or ii) there is at least one IoE device but none of them are activated. 
\end{definition}
When there is no IoE device in $\ncalH_L$, this face is inactive. We describe another extreme situation as shown in Fig.\ref{fig:outerenvelope}. When one or more IoE devices are randomly distributed in $\ncalH_L$ but there is no ES inside the outer envelope $\ncalE_{out}^L$, the devices located anywhere in $\ncalH_L$ can not be activated. Therefore, $\ncalH_L$ can not be activated as well. Considering these two cases, the probability $\P\{\ncalH_L \rm{\ is\ inactive}\}$ satisfies:
\begin{equation}
\begin{array}{@{}r@{}l}
    \P&\{\ncalH_L \rm{\ is\ inactive}\}\\
    &\geq e^{-\frac{3\sqrt{3}}{2}\lambda_r a_L^2}+(1-e^{-\frac{3\sqrt{3}}{2}\lambda_r a_L^2})e^{-\lambda_f S_{\ncalE_{out}^L}(r_r,r_f)},
\end{array}
\end{equation}
where $S_{\ncalE_{out}^L}(r_r,r_f)$ is the area of $\ncalE_{out}^L$.

% \begin{figure}
%     \centering
%     \includegraphics[width=1\columnwidth]{Figures/OuterEnvelope.pdf}
%     \caption{Illustration for the outer envelope in sub-critical case. If there is no energy station inside $\ncalE_{out}^L$, all IoE devices in the hexagon can not be activated.}
%     \label{fig:outerenvelope}
% \end{figure}

% The probability that $\ncalH_L$ is inactive satisfies:
% \begin{equation}
%     P(\ncalH\ is\ inactive)>P(No\ device\ in\ \ncalH)+P(At\ least\ one\ device\ in\ \ncalH)P(No\ ES\ in\ \ncalE_{out}),
% \end{equation}

When $\P\{\ncalH_L \rm{\ is\ inactive}\}>\frac{1}{2}$, there exists a sequence of connected inactive faces which form an inactive path. The inactive path which starts and ends at the same face is defined as an inactive circuit as shown in Fig.\ref{fig:ClosedCircuit}. We use $\ncalC(0)$ to represent the inactive circuit around the origin. Then, the number of active faces around the origin is finite, \ie $|K_{\ncalL_h^L}(0)|<\infty$. Because the side length of the hexagon $a_L=r_r$, the devices inside $\ncalC(0)$ can not be connected with those outside $\ncalC(0)$ directly or indirectly. Then, the number of devices connected with the device at the origin is finite, \ie $|K_{\ncalL_h^L}(0)|<\infty \rightarrow |K(0)|<\infty$. Based on this, we introduce the lower bound of critical ES density $\lambda_f^c$ in Lemma \ref{lem:lowerbound}.
\begin{lemma}[Lower bound of critical ES density]
    Mapping to the hexagonal lattice $\ncalL_h^L$, the lower bound of critical ES density is
\begin{equation}
    \lambda_f^L(\lambda_r,r_r,r_f)=\frac{1}{S_{\ncalE_{out}^L}(r_r,r_f)} \ln \frac{1-\exp(-\frac{3\sqrt{3}}{2}\lambda_r r_r^2)}{\frac{1}{2}-\exp(-\frac{3\sqrt{3}}{2}\lambda_r r_r^2)},
\label{lambdafl}
\end{equation}
where
\begin{equation}
    \lambda_r>\frac{2\ln 2}{3\sqrt{3}r_r^2},
\end{equation}
and
\begin{equation}
    S_{\ncalE_{out}^L}(r_r,r_f)=\frac{3\sqrt{3}}{2}r_r^2+6r_r r_f+\pi r_f^2.
\label{souterL}
\end{equation}
\label{lem:lowerbound}
\end{lemma}
\begin{IEEEproof}
See Appendix~\ref{app:lowerbound}.
\end{IEEEproof}
\begin{remark}
    Based on the lower bound in Lemma \ref{lem:lowerbound}, we obtain the sufficient condition for no percolation ($i.e.$, $\theta(\lambda_r,r_r,\lambda_f,r_f)=0$): i) $\lambda_r<\frac{2\ln 2}{3\sqrt{3}r_r^2}$ or ii) $\lambda_f<\lambda_f^L(\lambda_r,r_r,r_f), \lambda_r>\frac{2\ln 2}{3\sqrt{3}r_r^2}$. Because $0.768<\lambda_c(1)<3.37$, $\frac{2\ln{2}}{3\sqrt{3} r_r^2} \approx \frac{0.267}{r_r^2} < \frac{\lambda_c(1)}{r_r^2}$, the sufficient condition for no percolation can be updated as: i) $\lambda_r<\frac{\lambda_c(1)}{r_r^2}$ or ii) $\lambda_f<\lambda_f^L(\lambda_r,r_r,r_f), \lambda_r>\frac{\lambda_c(1)}{r_r^2}$. 
\label{rem:suflower}
\end{remark}
We can notice from (\ref{lambdafl}) that as the density of IoE device $\lambda_r$ approaches $\infty $, the lower bound for the critical density of ESs $\lambda_f^L$ approaches $\frac{\ln 2}{S_{\ncalE_{out}^L}}$. 

\subsection{Super-critical regime}
\label{subsec:sup}
\begin{figure}[htbp]
\centering
\subfigure[Active faces.]{
\begin{minipage}[t]{0.9\linewidth}
\centering 
\includegraphics[width=0.8\textwidth]{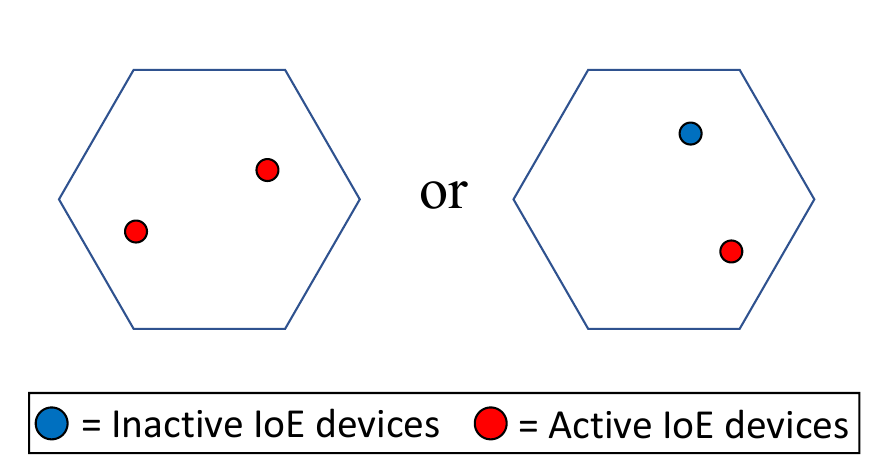}
\label{fig:OpenFace}
\end{minipage}}

\subfigure[Inner envelope.]{
\begin{minipage}[t]{0.9\linewidth}
\centering 
\includegraphics[width=1\textwidth]{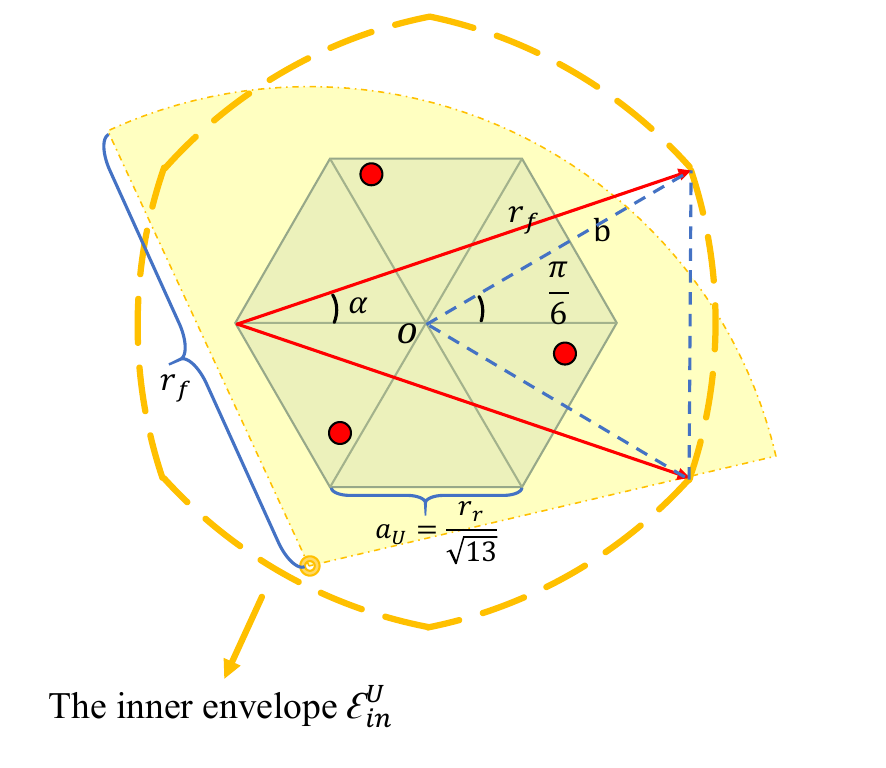}
\label{fig:innerenvelope}
\end{minipage}}

\subfigure[Face percolation.]{
\begin{minipage}[t]{1\linewidth}
\centering 
\includegraphics[width=1\textwidth]{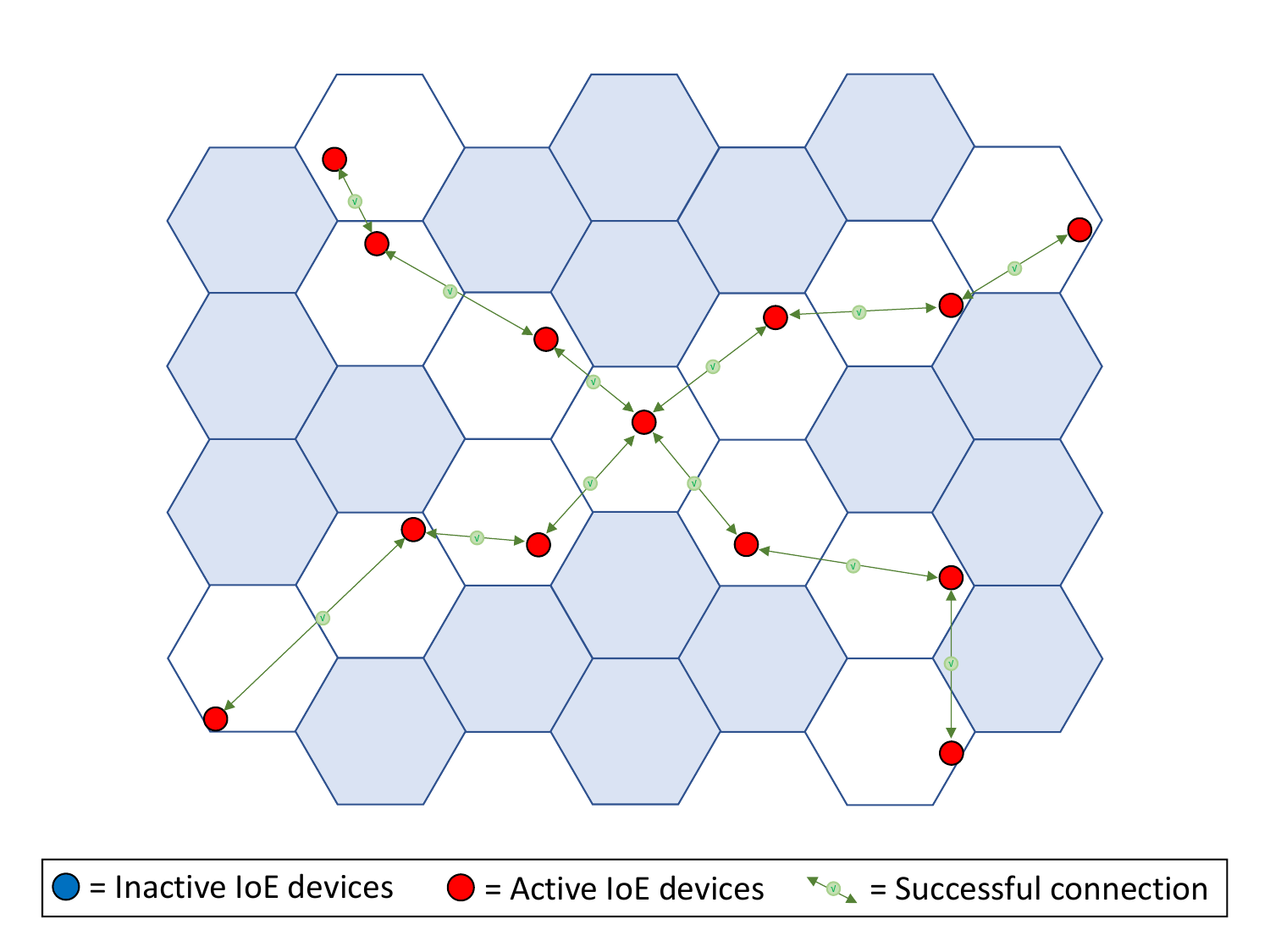}
\label{fig:FacePercolation}
\end{minipage}}
\caption{Illustration for the active faces, inner envelope, and face percolation. As shown in Fig.\ref{fig:OpenFace}, a hexagon with at least one active IoE device is considered an active face. Fig.\ref{fig:innerenvelope} is an extreme case we use to derive the upper bound. If there is an ES in the inner envelope, all IoE devices in the hexagon are activated. Fig.\ref{fig:FacePercolation} shows that face percolation happens when $\P\{\ncalH_U \rm{\ is\ active}\}>\frac{1}{2}$. Because $r_r=\sqrt{13}a_U$, the active IoE devices in two adjacent hexagons can communicate with each other, which helps the IoE network percolate at the same time.}
\end{figure}

This section proves that the probability of forming a giant connected component in the network is non-zero when ES density is dense enough. We also find the sufficient condition for non-zero percolation and derive the upper bound of critical ES density $\lambda_f^c$. Similar to the sub-critical regime, we analyze the super-critical regime by mapping the WC-RGG $G(V,E)$ to a hexagonal lattice, however, with different dimensions.\\ %\textcolor{red}{We also consider a special case where $r_r$ is the same as the maximum distance between devices in two adjacent hexagons, respectively.}\\
\textbf{Mapping to a Hexagonal Lattice in Super-critical Regime:} Let $\ncalL_h^U$ be a hexagonal lattice with a side length of $a_U=\frac{r_r}{\sqrt{13}}$ and let $\ncalH_U$ be a randomly selected hexagon (i.e., face) of $\ncalL_h^U$. In this case, $\sqrt{13}a_U=r_r$ is the maximum distance between any two IoE devices in two adjacent hexagons, hence, all IoE devices in any two adjacent hexagons are able to connect.
\begin{definition}[Active face in super-critical regime] A face $\ncalH_U$ is denoted as active if it contains at least one IoE device that lies within a WET zone of any ES. As shown in Fig.\ref{fig:OpenFace}, to make $\ncalH_U$ be an active face, the sufficient condition is: i) there is one or more IoE devices and ii) at least one of them is activated.     
\end{definition}

% \begin{figure}
%     \centering
%     \includegraphics[width=1\columnwidth]{Figures/Innerenvelope.pdf}
%     \caption{Illustration for the inner envelope in super-critical case. If there is at least one energy station inside $\ncalE_{in}^U$, all IoT devices in the hexagon are activated.}
%     \label{fig:innerenvelope}
% \end{figure}

To derive an upper bound expression for the critical ES density, we consider the situation in Fig.\ref{fig:innerenvelope}: there is at least one IoE device in $H_U$ and at least one ES inside the inner envelope $\ncalE_{in}^U$. In this case, the hexagon $\ncalH_U$ is active because there is at least one active IoE device in $H_U$. The probability $\P\{\ncalH_U \rm{\ is\ active}\}$ satisfies:
\begin{equation}
\begin{array}{@{}r@{}l}
    \displaystyle \P&\{\ncalH_U \rm{\ is\ active}\}\\
    &\geq(1-e^{-\frac{3\sqrt{3}}{2}\lambda_r a_U^2})(1-e^{-\lambda_f S_{\ncalE_{in}^U}(r_r,r_f)}),
\end{array}
\end{equation}
where $S_{\ncalE_{in}^U}(r_r,r_f)$ is the area of $\ncalE_{in}^U$.\\
\indent We also assume that there is an active IoE device at the origin. The face percolation happens in the hexagonal lattice when $\P\{\ncalH_U \rm{\ is\ active}\}>\frac{1}{2}$. An example of such a scenario is shown in Fig.\ref{fig:FacePercolation}. At the same time, the number of IoE devices connected with the device at the origin is infinite, \ie $|K_{\ncalL_h^U}(0)|=\infty \rightarrow|K(0)|=\infty$. Even if we overestimate the probability of an active face, it is available to analyze the super-critical regime. Based on this, we introduce the upper bound of critical ES density $\lambda_f^c$ in Lemma \ref{lem:upperbound}.
\begin{lemma}[Upper bound of critical ES density] Mapping to the hexagonal lattice $\ncalL_h^U$, the upper bound of critical ES density is 
\begin{equation}
    \lambda_f^U(\lambda_r,r_r,r_f)=\frac{1}{S_{\ncalE_{in}^U}(r_r,r_f)} \ln \frac{1-\exp(-\frac{3\sqrt{3}}{26}\lambda_r r_r^2)}{\frac{1}{2}-\exp(-\frac{3\sqrt{3}}{26}\lambda_r r_r^2)}, 
\label{lambdafu}
\end{equation}
where
\begin{equation}
    \lambda_r>\frac{26\ln 2}{3\sqrt{3}r_r^2},
\end{equation}
and
\begin{equation}
    S_{\ncalE_{in}^U}(r_r,r_f)=\frac{3\sqrt{3}}{2}b^2+6\alpha r_f^2-3 r_f^2 \sin 2\alpha.
\label{sinnerU}
\end{equation}
with $b=\sqrt{r_f^2-\frac{r_r^2}{52}}-\frac{\sqrt{39}r_r}{26}$ and $\alpha=\arcsin \frac{b}{2r_f}$.
\label{lem:upperbound}
\end{lemma}
\begin{IEEEproof}
    See Appendix~\ref{app:upperbound}
\end{IEEEproof}
\begin{remark}
    Based on the upper bound in Lemma \ref{lem:upperbound}, we obtain the sufficient condition for non-zero percolation  ($i.e.$, $\theta(\lambda_r,r_r,\lambda_f,r_f)>0$): i) $\lambda_r>\frac{26\ln 2}{3\sqrt{3}r_r^2}$ and ii) $\lambda_f>\lambda_f^U(\lambda_r,r_r,r_f)$. Because $\frac{26\ln 2}{3\sqrt{3}r_r^2}\approx \frac{3.46}{r_r^2}>\frac{\lambda_c(1)}{r_r^2}$, this condition is sufficient enough. 
\label{rem:sufupper}
\end{remark}
We can notice from (\ref{lambdafu}) that as the density of IoE device $\lambda_r$ approaches $\infty$, the upper bound of the critical density of ESs $\lambda_f^U$ approaches $\frac{\ln 2}{S_{\ncalE_{in}^U}}$.

\subsection{Critical ES density function}
\label{subsec:cri}
\indent We have introduced the percolation probability in (\ref{perpro}). To analyze the network-wide D2D connectivity in RF-powered IoE networks, it is important to find the critical ES density $\lambda_f^c$ for the phase transition of the percolation probability. The phase transition is defined in Theorem \ref{theo:perpro}.
%  about the density of energy stations
\begin{theorem}[Phase transition] Let $\theta(\lambda_r,r_r,\lambda_f,r_f)$ denote the percolation probability in the RGG $G(V,E)$, $\forall \lambda_r>\frac{\lambda_c(1)}{r_r^2}$, there exists a critical value $\lambda_f^c<\infty$ for the density of ESs such that
\begin{equation}
    \begin{array}{ll}
       \theta(\lambda_r,r_r,\lambda_f,r_f)=0  & for\,\, \lambda_f < \lambda_f^c  \\
       \theta(\lambda_r,r_r,\lambda_f,r_f)>0  & for\,\, \lambda_f > \lambda_f^c  \\
    \end{array}.
\end{equation}
\label{theo:perpro}
\end{theorem}
\begin{IEEEproof}
    We first prove that the percolation probability $\theta(\lambda_r,r_r,\lambda_f,r_f)$ is a non-decreasing function of $\lambda_f$ when $\lambda_r>\frac{\lambda_c(1)}{r_r^2}$. Consider two sets of ESs $\Psi_1$ and $\Psi_2$ with densities $\lambda_{f_1}$ and $\lambda_{f_2}$, respectively, where $\lambda_{f_1}<\lambda_{f_2}$. Since $\Psi_1$ and $\Psi_2$ are both PPPs, $\Psi_1$ can be constructed by thinning $\Psi_2$ with probability $\frac{\lambda_{f_1}}{\lambda_{f_2}}$, thus $\Psi_1 \subseteq \Psi_2$. The set of active IoE devices is defined as $V_l=\{x_i\in\Psi: \min\limits_{y_k\in\Psi_l}\|x_i-y_k\|\leq r_f\}$ for $l=\{1,2\}$. Hence,  $|V_1\cap \ncalA|\leq|V_2\cap \ncalA|$ for any $\ncalA\subset \R^2$. Considering the connected component with the IoE device at origin, obviously $K_1(0)\subseteq K_2(0)$ and $|K_1(0)|\leq|K_2(0)|$. Therefore, $0<\lambda_{f_1}<\lambda_{f_2}$ indicates $\theta(\lambda_r,r_r,\lambda_{f_1},r_f)\leq\theta(\lambda_r,r_r,\lambda_{f_2},r_f)$. That means the percolation probability is a non-decreasing function of $\lambda_f$.\\% Based on Kolmogorov's zero-one law, we prove that
    \indent In Remark \ref{rem:suflower} and Remark \ref{rem:sufupper}, we derive the sufficient conditions for no percolation and non-zero percolation. We know that $\theta(\lambda_r,r_r,\lambda_f,r_f)=0$ for $\lambda_f<\lambda_f^L$ and $\theta(\lambda_r,r_r,\lambda_f,r_f)>0$ for $\lambda_f>\lambda_f^U$. As shown in (\ref{mperpro}), when $\lambda_f=\infty$, the percolation only depends on the density of IoE devices, \ie $\theta(\lambda_r,r_r,\infty,r_f)=\theta(\lambda_r,r_r)$. The sufficient condition for $\theta(\lambda_r,r_r)>0$ is $\lambda_r>\frac{\lambda_c(1)}{r_r^2}$, which means percolation probability is not always zero when increasing the value of $\lambda_f$, $\forall \lambda_r>\frac{\lambda_c(1)}{r_r^2}$. Since $\theta(\lambda_r,r_r,\lambda_f,r_f)$ is a non-decreasing function of $\lambda_f$, there should be a critical value $\lambda_f^c$ which exhibits the phase transition shown in Theorem \ref{theo:perpro}.
\end{IEEEproof}
To analyze the relationship between $\lambda_f^c$ and $\lambda_r$, we introduce Lemma \ref{lem:criticaldensity}.
\begin{lemma}[Critical ES density function] The critical density of ESs is defined as a function written as $\lambda_f^c(\lambda_r,r_r,r_f)$, which is a non-increasing function of $\lambda_r$ satisfying:
\begin{equation}
        \begin{array}{ll}
       \theta(\lambda_r,r_r,\lambda_f,r_f)=0  & for\,\, \lambda_f < \lambda_f^c(\lambda_r,r_r,r_f)  \\
       \theta(\lambda_r,r_r,\lambda_f,r_f)>0  & for\,\, \lambda_f > \lambda_f^c(\lambda_r,r_r,r_f)  \\
    \end{array},
\end{equation}
where
\begin{equation}
    \lambda_r>\frac{\lambda_c(1)}{r_r^2}.
\end{equation}
\label{lem:criticaldensity}
\end{lemma}
\begin{IEEEproof}
    When $\lambda_f=\infty$, the percolation probability is updated as $\theta(\lambda_r,r_r)$. As a common conclusion in percolation theory, it is a non-decreasing function of $\lambda_r$. When $\lambda_f<\infty$ and $\lambda_r<\frac{\lambda_c(1)}{r_r^2}$, $\theta(\lambda_r,r_r,\lambda_f,r_f)=0$. To prove Lemma \ref{lem:criticaldensity}, we consider two sets of IoE devices $\Phi_1$ and $\Phi_2$ with densities $\lambda_{r_1}$ and $\lambda_{r_2}$, respectively, where $\frac{\lambda_c(1)}{r_r^2}<\lambda_{r_1}\leq\lambda_{r_2}$. Since $\Phi_1$ and $\Phi_2$ are both PPPs,  $\Phi_1$ can be constructed by thinning $\Phi_2$ with probability $\frac{\lambda_{r_1}}{\lambda_{r_2}}$ and $\Phi_1\subseteq\Phi_2$. $K_1'(0)$ and $K_2'(0)$ are the connected component from active IoE devices in $\Phi_1$ and $\Phi_2$, respectively. Therefore, $K_1'(0)\subseteq K_2'(0)$ and $|K_1'(0)|\leq|K_2'(0)|$. Similar to the proof of Theorem \ref{theo:perpro}, percolation probability is a non-decreasing function of $\lambda_r$. We define $\lambda_{f_1}^c=\lambda_f^c(\lambda_{r_1},r_r,r_f)$ and $\lambda_{f_2}^c=\lambda_f^c(\lambda_{r_2},r_r,r_f)$. Then $\theta(\lambda_{r_2},r_r,\lambda_{f_1}^c,r_f)\geq\theta(\lambda_{r_1},r_r,\lambda_{f_1}^c,r_f)=0$, while $\theta(\lambda_{r_2},r_r,\lambda_{f_2}^c,r_f)=0$. Therefore, $\theta(\lambda_{r_2},r_r,\lambda_{f_1}^c,r_f)\geq\theta(\lambda_{r_2},r_r,\lambda_{f_2}^c,r_f)=0$ for $\frac{\lambda_c(1)}{r_r^2}<\lambda_{r_1}\leq\lambda_{r_2}$.\\ 
    \indent  When $\theta(\lambda_{r_2},r_r,\lambda_{f_1}^c,r_f)=\theta(\lambda_{r_2},r_r,\lambda_{f_2}^c,r_f)=0$, $\lambda_{f_1}^c\leq\lambda_{f_2}^c$ because percolation probability is a non-decreasing function of $\lambda_f$ and $\lambda_{f_2}^c$ is the critical ES density of phase transition when $\lambda_r=\lambda_{r_2}$. If we assume that $\lambda_{f_1}^c<\lambda_{f_2}^c$ for $\lambda_{r_1}\leq\lambda_{r_2}$, $\theta(\lambda_{r_2},r_r,\lambda_{f_2}^c,r_f)=0$ is impossible because $\theta(\lambda_{r_2},r_r,\lambda_{f_2}^c,r_f)\geq\theta(\lambda_{r_1},r_r,\lambda_{f_2}^c,r_f)>\theta(\lambda_{r_1},r_r,\lambda_{f_1}^c,r_f)=0$. So when $\frac{\lambda_c(1)}{r_r^2}<\lambda_{r_1}\leq\lambda_{r_2}$, if $\theta(\lambda_{r_2},r_r,\lambda_{f_1}^c,r_f)=\theta(\lambda_{r_2},r_r,\lambda_{f_2}^c,r_f)=0$, $\lambda_{f_1}^c=\lambda_{f_2}^c$. This means that for the same value of $\lambda_r$, there is only a unique critical ES density, \ie $\lambda_{r_1}=\lambda_{r_2}$ implies $\lambda_f^c(\lambda_{r_1},r_r,r_f)=\lambda_f^c(\lambda_{r_2},r_r,r_f)$.\\
    \indent When $\theta(\lambda_{r_2},r_r,\lambda_{f_1}^c,r_f)>\theta(\lambda_{r_2},r_r,\lambda_{f_2}^c,r_f)=0$, because percolation probability is a non-decreasing function of $\lambda_r$, $\lambda_{r_2}>\lambda_{r_1}$. At the same time, $\lambda_{f_2}^c=\lambda_f^c(\lambda_{r_2},r_r,r_f)<\lambda_{f_1}^c$ because the percolation probability is also a non-decreasing function of $\lambda_f$ and $\lambda_f^c$ is the critical ES density. Thus, $\lambda_{r_1}<\lambda_{r_2}$ implies $\lambda_f^c(\lambda_{r_1},r_r,r_f)>\lambda_f^c(\lambda_{r_2},r_r,r_f)$.\\
    \indent In conclusion, $\lambda_{r_1}\leq\lambda_{r_2}$ implies $\lambda_f^c(\lambda_{r_1},r_r,r_f)\geq\lambda_f^c(\lambda_{r_2},r_r,r_f)$, and critical ES density is a non-decreasing function of $\lambda_r$.%, \ie $\lambda_f^c(\lambda_r,r_r,r_f)$ is a non-increasing function of $\lambda_r$.
\end{IEEEproof}

\section{Approximations}
In the previous section, we discuss the sub-critical and super-critical regimes and complete the proof of phase transition of percolation probability in RF-powered IoE networks under an EaaS platform. We define the critical ES density as a function of IoE device density. In this section, we use some classic tools in graph theory, such as the inner-city model and Gilbert disk models, to introduce tight approximations for critical ES density function, which are important to estimate the ESs deployment CAPEX to offer a successful EaaS.
\subsection{Inner City Model}
When the density of IoE devices is not very large, the distribution of active IoE devices follows a special model denoted as the inner-city model, which is introduced in \cite{haenggi2012stochastic}. The inner-city model is generated from the complementary process of the Poisson hole process. In particular, the inner-city model is a Cox process with the intensity field $\zeta(x)=\lambda_r 1_{\Xi_f(x)}$, where $\Xi_f\triangleq\bigcup\{\yk\in\Psi:\textbf{b}(\yk,r_f)\}$ is the union of all disks centered at ESs with radius $r_f$. We consider three steps: \textsl{i) generate the IoE devices using a PPP $\Phi$ with density $\lambda_r$, ii) generate the ESs using another PPP $\Psi$ with density $\lambda_f$, and iii) choose the IoE devices located inside the WET zones of ESs, which are the disks centered at the ESs $\textbf{b}(\yk,r_f)$ where $\yk\in\Psi$}. \\
\indent In \cite{haenggi2012stochastic} and \cite{7557010}, the Poisson hole process can be approximated using a homogeneous PPP with the same density, especially when the hole is small. As the complementary process of the Poisson hole process, we propose to analyze the inner-city model in a similar way. On the contrary, the inner-city model can be approximated well using a homogeneous PPP with the same density when $r_f$ is large because as many points outside the hole as possible are retained. We approximate this point process of active IoE devices as a PPP constructed by thinning the PPP $\Phi$ with probability $1-e^{-\lambda_f\pi r_f^2}$. The density of active IoE devices can be approximated using the product of the density of IoE devices and the probability that one IoE device is activated:
\begin{equation}
    \hat{\lambda}_r\approx\lambda_r (1-e^{-\lambda_f\pi r_f^2}).
\label{lambdarhat}
\end{equation}
\indent In the considered system model, two IoE devices can be connected when the distance between them is less than $r_r$. After we use the PPP with density $\hat{\lambda}_r$ to model the distribution of active IoE devices, their connectivity can be described as a Gilbert disk model $D(\hat{\lambda}_r,\frac{r_r}{2})$. For non-zero percolation, $\hat{\lambda}_r$ should satisfy:
\begin{equation}
    \hat{\lambda}_r>\frac{\lambda_c(1)}{r_r^2}.
\label{firstcritical}
\end{equation}
\begin{lemma}[Approximation of critical ES density using the inner-city model and Gilbert disk model]
    Using the inner-city model and Gilbert disk model $D(\hat{\lambda}_r,\frac{r_r}{2})$, the approximation of critical ES density is written as:
\begin{equation}
    \lambda_f^{IC}(\lambda_r,r_r,r_f)=-\frac{1}{\pi r_f^2} \ln (1-\frac{\lambda_c(1)}{\lambda_r r_r^2}).
\end{equation}
\label{firstapproximation}
\end{lemma}
\begin{IEEEproof}
When $\lambda_r$ is fixed, we substitute (\ref{lambdarhat}) in (\ref{firstcritical}), $\lambda_f$ should satisfy
\begin{equation}
    \lambda_f>-\frac{1}{\pi r_f^2} \ln (1-\frac{\lambda_c(1)}{\lambda_r r_r^2}).
\end{equation}
\end{IEEEproof}
\begin{remark}
    In order to make the distribution of active IoE devices sufficiently dispersed and uniform, this approximation is more suitable when $\lambda_r$ is small. When $\lambda_r$ is very large, the active IoE devices are dense only around the ESs, but not uniform in the whole space. The value of the lower bound approaches a fixed value, but the value of $\lambda_f^{IC}$ is 0, which is not matching enough. 
\end{remark}
To approximate the critical ES density when $\lambda_r$ is large, we build a simple Gilbert disk model to analyze it. 
\subsection{Simple Gilbert Disk}
\begin{figure}
    \centering
    \includegraphics[width=0.85\columnwidth]{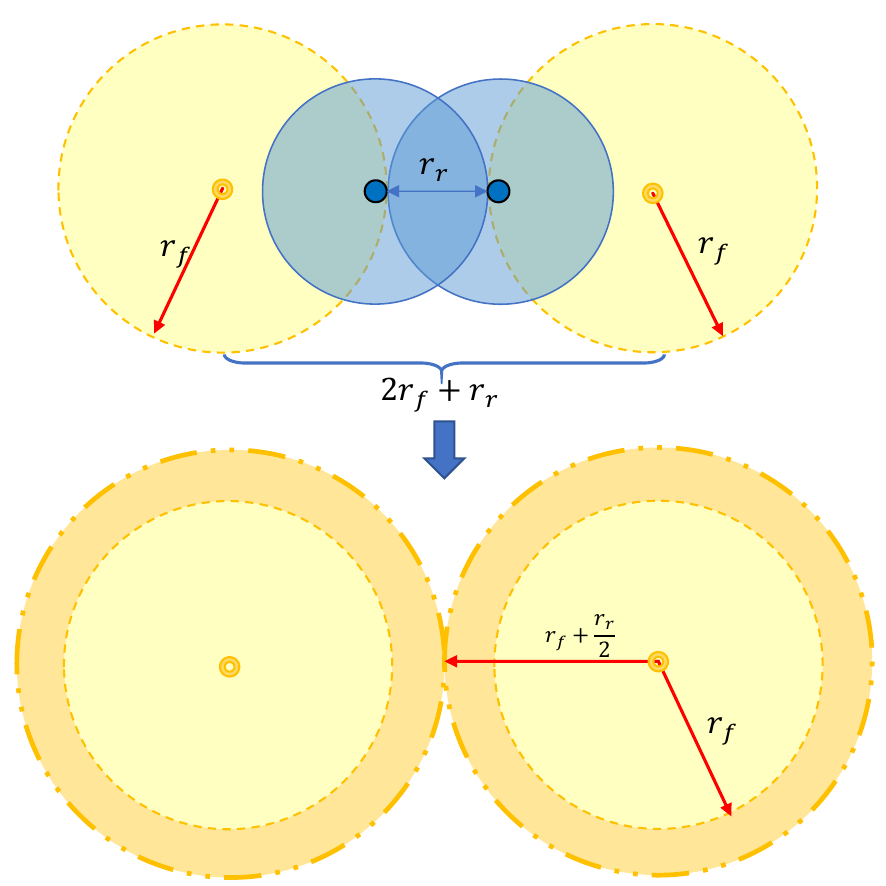}
    \caption{Illustration for the approximation based on a simple Gilbert disk model. When the density of IoE devices is very large the necessary condition for connecting IoE devices in two adjacent WET areas is to make the distance between these two adjacent ESs (the centers of WET zones) at most $2r_f+r_r$, where the distance between the two points at the critical position (marked in blue) is exactly the maximum communication range $r_r$.}
    \label{fig:secondappro}
\end{figure}

%When the density of IoE devices is very large, the IoE devices inside the WET zone of the same ES can be connected together. But for non-zero percolation in the whole large-scale network, we consider the critical case as shown in Fig.\ref{fig:secondappro}. The connectivity of active IoE devices can be described as another Gilbert disk model $D(\lambda_f,\frac{2r_f+r_r}{2})$, which is named as a `simple Gilbert disk model' in Lemma \ref{lem:secondapproximation} because we assume that $\lambda_r=\infty$ and the approximation of critical ES density only depends on $r_f$ and $r_r$. 
When the density of IoE devices is very large, we can safely assume that IoE devices within similar WET zones are forming connected components. However, for connected components of different WET zones to establish a connection and achieve non-zero percolation, we need to ensure the percolation of the Gilbert disk model $D(\lambda_f,\frac{2r_f+r_r}{2})$ shown in Fig.\ref{fig:secondappro} (referred to as ``simple Gilbert disk model" in Lemma \ref{lem:secondapproximation}.

\begin{lemma}[Approximation of critical ES density using a simple Gilbert disk model]
    Using a simple Gilbert disk model $D(\lambda_f,\frac{2r_f+r_r}{2})$, the approximation of critical ES density is written as:
\begin{equation}
    \lambda_f^{GD}(\lambda_r,r_r,r_f)=\frac{\lambda_c(1)}{(2r_f+r_r)^2}.
\label{lamfGD}
\end{equation}
\label{lem:secondapproximation}
\end{lemma}
\begin{IEEEproof}
As shown in Fig.\ref{fig:secondappro}, when $\lambda_r$ is fixed, the distance between two ESs is $2r_f+r_r$. For non-zero percolation in Gilbert disk model $D(\lambda_f,\frac{2r_f+r_r}{2})$, $\lambda_f$ should satisfy
\begin{equation}
    \lambda_f>\frac{\lambda_c(1)}{(2r_f+r_r)^2}.
\end{equation}
\end{IEEEproof}
\begin{remark}
    This model considers any ES and its nearby IoE devices as a whole. So this approximation is more suitable when $\lambda_r$ is large enough. When $\lambda_r$ is small, the value of this approximation is much less than the exact critical ES density, which is also not valid enough.
\end{remark}
Though $\lambda_f^{GD}$ can approximate the critical ES density when $\lambda_r$ is large, it can not replace $\lambda_f^{IC}$ when $\lambda_r$ is small. Comparing the lower bound and upper bound, we propose a novel idea to combine the advantages of $\lambda_f^{IC}$ and $\lambda_f^{GD}$ without losing tractability.
\subsection{Approximation from imitating bounds}
The approximations $\lambda_f^{IC}$ and $\lambda_f^{GD}$ are suitable in different ranges of $\lambda_r$. In this part, we try to combine their advantages together. Similar to the lower bound and the upper bound, we use an expression like them to approximate the critical ES density in Lemma \ref{lem:approximationbound}.
\begin{lemma}
The approximation of critical ES density is written as:
\begin{equation}
    \lambda_f^*(\lambda_r,r_r,r_f)=\frac{\lambda_c(1)/ \ln 2}{(2r_f+r_r)^2}\ln \frac{1-\exp(-\frac{\ln 2}{\lambda_c(1)}\lambda_r r_r^2)}{\frac{1}{2}-\exp(-\frac{\ln 2}{\lambda_c(1)}\lambda_r r_r^2)},
\label{lamf*}
\end{equation}
where
\begin{equation}
    \lambda_c(1)=\frac{4\ln 2}{\pi}\approx 0.883.
\end{equation}
% where $k=\sqrt{\frac{9\ln 2}{\pi \lambda_c(1)}}-1\approx 0.174$ and $\gamma=\frac{r_r}{r_f}$.
\label{lem:approximationbound}
\end{lemma}

\begin{IEEEproof}
    See Appendix~\ref{app:approximationbound}.
\end{IEEEproof}
\indent This approximation is a combination of the approximations $\lambda_f^{IC}$ and $\lambda_f^{GD}$. It is obtained by imitating the expressions of upper and lower bounds, which are compared in Table.\ref{tab:comparison}. We justify this approximate expression and propose a conjecture about the value of $\lambda_c(1)$ in Appendix~\ref{app:approximationbound}, which is in agreement with the upper bound of critical density for 2D site percolation in \cite{kirkwood1983percolation} and tight lower bound on the critical density for 2D Poisson RGGs in \cite{4557082}.

\begin{table*}[htbp]\caption{Comparison between Sub-critical regime and Super-critical regime}
\centering
\begin{center}
\resizebox{\textwidth}{!}{
\renewcommand{\arraystretch}{1.5}%1.4
    \begin{tabular}{ {c} | {c}|  {c}}
    \hline
        \hline
    \textbf{Regime} & \textbf{Bound} &  \textbf{Threshold} \\ \hline
    Sub-critical & Lower bound $\lambda_f^L(\lambda_r,r_r,r_f)=\displaystyle\frac{1}{S_{\ncalE_{out}^L}(r_r,r_f)} \ln \frac{1-\exp(-\frac{3\sqrt{3}}{2}\lambda_r r_r^2)}{\frac{1}{2}-\exp(-\frac{3\sqrt{3}}{2}\lambda_r r_r^2)}$ & $\lambda_r=\frac{2\ln 2}{3\sqrt{3}r_r^2}\approx \frac{0.267}{r_r^2}$  \\ \hline
    Super-critical & Upper bound $\lambda_f^U(\lambda_r,r_r,r_f)=\displaystyle\frac{1}{S_{\ncalE_{in}^U}(r_r,r_f)} \ln \frac{1-\exp(-\frac{3\sqrt{3}}{26}\lambda_r r_r^2)}{\frac{1}{2}-\exp(-\frac{3\sqrt{3}}{26}\lambda_r r_r^2)}$ & $\lambda_r=\frac{26\ln 2}{3\sqrt{3}r_r^2}\approx \frac{3.468}{r_r^2}$  \\ \hline
    \hline
    \textbf{Regime} & \textbf{Area of Envelope} &  \textbf{Limit Value of Bound} \\ \hline
    Sub-critical & $S_{\ncalE_{out}^L}(r_r,r_f)=\frac{3\sqrt{3}}{2}r_r^2+6r_r r_f+\pi r_f^2$ & $\ln 2/S_{\ncalE_{out}^L}(r_r,r_f)$ \\ \hline
    Super-critical & $S_{\ncalE_{in}^U}(r_r,r_f)=\frac{3\sqrt{3}}{2}b^2+6\alpha r_f^2-3r_f^2\sin 2\alpha$, where $b=\sqrt{r_f^2-\frac{1}{52}r_r^2}-\frac{\sqrt{39}}{26}r_r$ and $\alpha=\arcsin \frac{b}{2r_f}$ & $\ln 2 / S_{\ncalE_{in}^U}(r_r,r_f)$ \\ \hline
    \hline
    \end{tabular}
    }
\end{center}
\label{tab:comparison}
%\vspace{-8mm}
\end{table*}

%\in\Big(\pi(r_f+\frac{\sqrt{3}}{2}r_r)^2,\pi(r_f+r_r)^2\Big)
%\in\Big(\pi(r_f-\frac{1}{\sqrt{13}}r_r)^2,\pi b^2\Big)
\section{Simulation results and discussion}
\begin{figure}
    \centering
    \includegraphics[width=1\columnwidth]{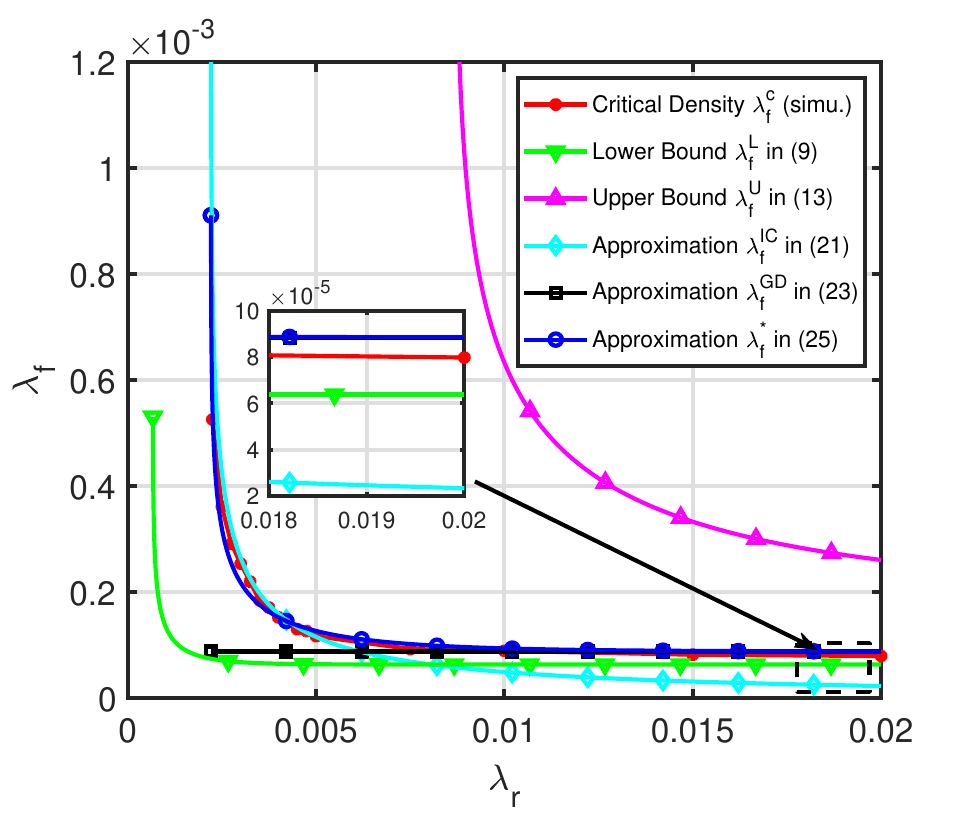}
    \caption{Comparison between curves of bounds, approximations, and simulation results of the critical ES density. As the density of IoE devices increases, the bounds, approximations and simulated critical ES density do not increase.}
    \label{fig:lambdaf_lambdar}
\end{figure}

\begin{figure}
    \centering
    \includegraphics[width=1\columnwidth]{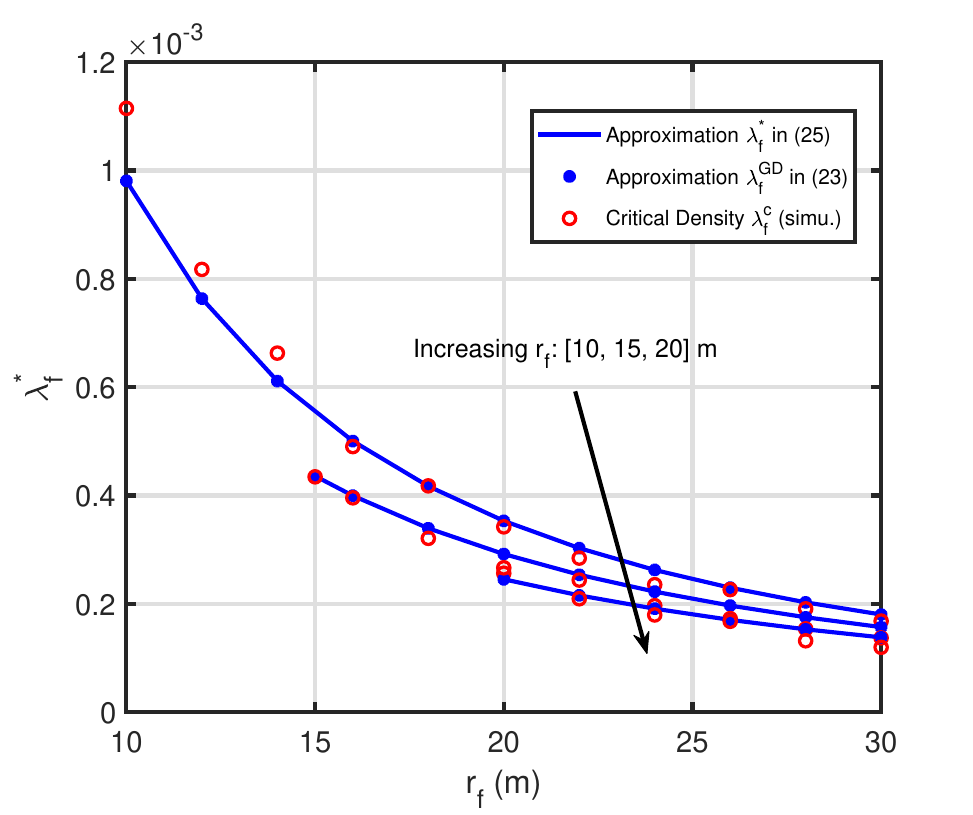}
    \caption{Comparison between theoretical results and approximation of critical ES density when $\lambda_r\gg \frac{\lambda_c(1)}{r_r^2}$. The derived approximation $\lambda_f^{*}$ in (\ref{lamf*}) approaches $\lambda_{f}^{GD}$ in (\ref{lamfGD}) when IoE devices are dense.}
    \label{fig:lambdaf_rf}
\end{figure}

\begin{figure}
    \centering
    \includegraphics[width=1\columnwidth]{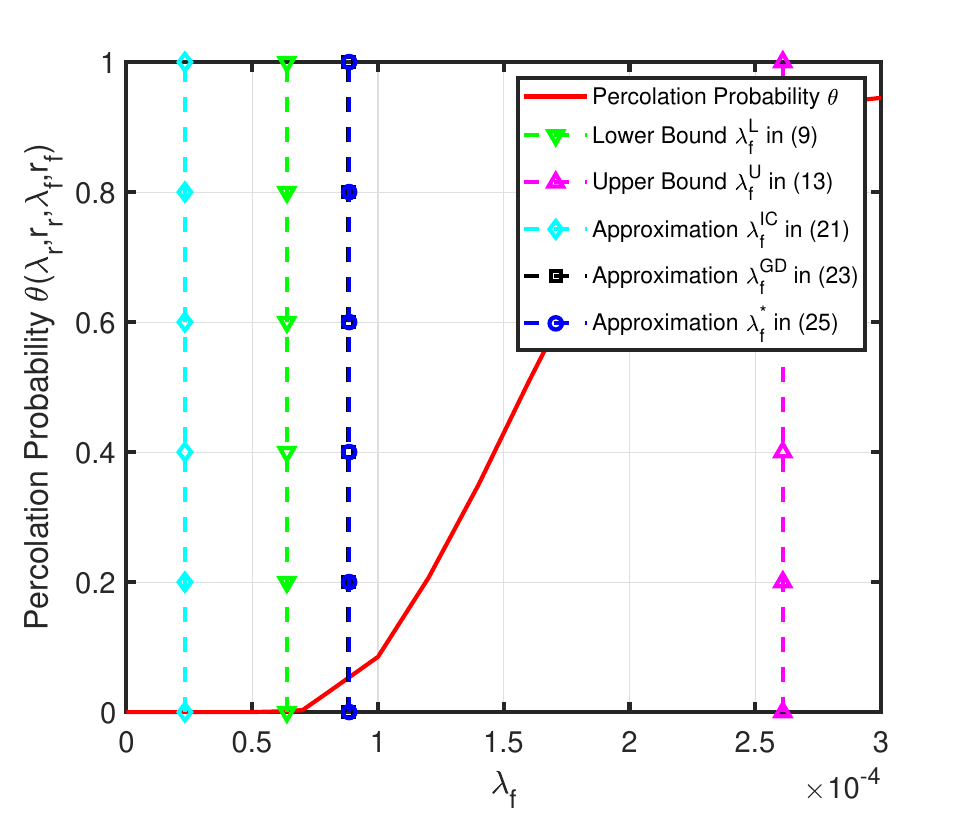}
    \caption{The simulation result of percolation probability $\theta(\lambda_r,r_r,\lambda_f,r_f)$ in $\lambda_f$ when $\lambda_r\gg \frac{\lambda_c(1)}{r_r^2}$. The phase transition of percolation probability happens between the derived bounds and approximations $\lambda_{f}^{GD}$ and $\lambda_{f}^{*}$ work.}
    \label{fig:perpro}
\end{figure}

%\indent \textcolor{cyan}{In this paper, we simulate in a 4000\,{\rm m} × 4000\,{\rm m} rectangular area, and do 10,000 iterations in each iteration. If we have a giant component horizontally or vertically, we think the graph of active IoT devices percolates. We record the value of critical ES density and calculate the expectation, in order to verify that our approximations and actual critical ES density match.} 
\indent In this section we show simulation results for the percolation probability of the considered EaaS system compared to the derived bounds. We consider a 4000\,{\rm m} × 4000\,{\rm m} rectangular area and conduct Monte-Carlo simulations with 10,000 iterations. For each iteration, we consider that percolation occurs if we either have a giant connected component horizontally or vertically, where the distance between farthest two points is much larger than the D2D communication range and close to the simulation boundaries. We compute the average of critical ES density over all iterations. We put the bounds and approximations together in Fig.\ref{fig:lambdaf_lambdar}. Based on \cite{ahamad2021interference,wang2019social,salim2022rf,zhao2016social}, it is common to assume that the maximum D2D communication distance $r_r$ is smaller than 40 {\rm m}. The maximum WET range $r_f$ can be much larger than $r_r$ \cite{lu2014wireless}. In the proposed system, the WET range of each ES is considered larger than the D2D communication range to ensure a sufficiently large energy supply range. Therefore, we set $r_r=20\,{\rm m}$ and $r_f=40\,{\rm m}$. The simulation result is a curve between the lower bound $\lambda_f^L$ and the upper bound $\lambda_f^U$. The value of the upper bound is always the highest, on the contrary, the value of the lower bound is always the lowest. $\lambda_f^{IC}$ and $\lambda_f^{GD}$ are suitable approximations in different range of $\lambda_r$. At the same time, $\lambda_f^*$ is an approximation for all $\lambda_r>\frac{\lambda_c(1)}{r_r^2}$. When $\lambda_r$ is small, the approximations $\lambda_f^{IC}$ and $\lambda_f^*$ are similar, which are located between the upper and lower bound. The value of $\lambda_f^{GD}$ is much less than the simulation. When $\lambda_r$ is large enough, the simulation $\lambda_f^c$, approximations $\lambda_f^{GD}$ and $\lambda_f^*$ are matching, while $\lambda_f^{IC}$ is too small. \\
\indent We also compare the approximation and simulation result in Fig.\ref{fig:lambdaf_rf} when $\lambda_r\gg\frac{\lambda_c(1)}{r_r^2}$ which are matching with a small gap. When we choose different values of $r_r$, the critical density decreases when $r_f$ increases. \\
\indent In Fig.\ref{fig:perpro}, we plot the curve of the simulation result of percolation probability $\theta(\lambda_r,r_r,\lambda_f,r_f)$ in $\lambda_f$. We set $r_r=20\,{\rm m}$, $r_f=40\,{\rm m}$, and $\lambda_r\gg \frac{\lambda_c(1)}{r_r^2}$. The percolation probability is a non-decreasing function of $\lambda_f$. When $\lambda_f$ is less than the lower bound $\lambda_f^L$, the percolation probability is 0. On the contrary, when $\lambda_f$ is larger than the upper bound $\lambda_f^U$, the percolation probability is non-zero. The approximations $\lambda_f^*$ and $\lambda_f^{GD}$ of the critical ES density are located between the lower bound $\lambda_f^L$ and the upper bound $\lambda_f^U$.\\
\indent To show the CAPEX gains, we can contrast the proposed percolation theory driven deployment against a full-coverage driven deployment. Recall that the critical ES density obtained by our percolation driven model aims to minimize the required density of ESs for full network connectivity. Alternatively, if the density of IoE devices is $0.01$ and ESs are deployed following a square lattice, the density of ESs for full-coverage is $\frac{1}{2 r_f^2}\approx 3.125\times 10^{-4}\,{\rm ESs/m^2}$, which is $2.26$ times larger than the critical density obtained by our approximation $\frac{\lambda_c(1)}{(2r_r+r_f)^2} \approx 1.378\times 10^{-4}\,{\rm ESs/m^2}$. Therefore, in the considered rectangular area of $4000\,{\rm m}\times 4000\,{\rm m}$ dimensions, our percolation driven model can help reduce the cost of deploying more than 2700 ESs. However, the CAPEX saving of our percolation driven model comes at the cost of spatial power outages, which restricts the deployment of the IoE devices to utilize the offered EaaS. In particular, our percolation theoretic deployment reveals the minimum density of ESs that maintains large-scale connectivity of RF-power activated IoE devices but does not maintain 100\% coverage of the spatial area. Based on the spatial randomness of the Gilbert disk model, the probability that an arbitrary location is not covered by the ESs is $1-e^{\lambda_f \pi r_f^2}$, which translates to approximately 0.5 for $\lambda_r=0.01$ and $\lambda_f=\lambda_f^c=\frac{\lambda_c(1)}{(2r_r+r_f)^2}$. More comprehensive spatial coverage for the IoE devices would either require a higher density of ESs or a special arrangement (e.g., via batteries) for devices not covered by ESs. \\
\indent It is worth noting that the theoretical analysis and simulations provided in this paper are based on an idealized system model. In realistic applications, the WET zones of ESs and D2D communication range can be limited by many factors, such as weather, terrain, buildings, vehicles, large metal machines, etc. Therefore, the WET zones of ESs are not always as regular as circular areas, and the critical ES density we obtain (\ie the approximations $\lambda_f^*$) becomes a lower bound of the actual ES density requirement. For realistic deployments, the lower bound $\lambda_f^*$ serves as a guideline to estimate the actual ES density requirement to realize the large-scale connectivity first and reduce the potential CAPEX (\ie redundant equipment manufacturing costs). 
It is also worth noting that the considered ESs themselves are assumed to have perpetual energy sources by being connected to the electricity grid. However, for rural and remote areas, connecting every ES to the electricity grid may be infeasible. Therefore, the ESs may depend on random energy sources through solar panels or wind turbines. At the same time, the distribution of IoE devices can be scattered. Therefore, the large-scale multi-hop connectivity in rural and remote areas still needs to be further discussed.\\
\indent The comparative analysis in the context of this paper shall mainly focus on comparing the WET EaaS platform for IoE against the classical battery-powered IoE. The performance of the battery-powered IoE can be considered as an upper bound for the WET EaaS-powered IoE, in which all IoE devices are activated by the electricity grid. However, it is significantly important to highlight the fact that battery-powered IoE can be placed in hard-to-reach or remote locations, making battery replacement/recharging a complicated process. In addition, the disposal of such batteries adds another level of complexity, especially when it needs to be done in an environment-friendly manner. All these aspects reflect the unique advantages of RF-powered IoE devices, which need to be included when comparing EaaS for IoE with battery-powered IoE. However, unlike connectivity and performance, which can be quantified through percolation theory as shown in this paper, the other aspects concerning the battery replacement/recharging difficulties and their disposal complexities are harder to quantify without more concrete operational data. In addition, to better investigate realistic IoE networks, it is expected to expand this EaaS platform to the case with heterogeneous IoE networks and heterogeneous ES networks.

\section{Conclusions}
% \indent In energy harvesting wireless networks, the connectivity between IoT devices not only depends on the density and communication range of these devices, but also is affected by the number of energy stations and the scope of energy supply. In order to ensure that enough IoT devices are activated and multi-hop connections are achieved, energy stations are randomly deployed in the entire space to support the circuit operation of their adjacent IoT devices. Therefore, we build the system model where IoT devices and ESs follow different and independent PPPs. \\
\indent This paper characterizes large-scale D2D connectivity of RF-powered IoE devices operating under EaaS platform, where energy providers deploy ESs that offer WET to power battery-less IoE devices. In this context, we develop a novel WET-aware connectivity random geometric graph assuming that the ESs and the D2D devices follow independent PPPs. Using percolation theory, we prove the existence of a finite critical ES density above which the WC-RGG operates in the super-critical regime, which implies network-wide large-scale D2D connectivity for the RF-powered IoE network. We show that the critical ES density decreases when the density of IoE devices increases. We also obtain approximations for the critical ES density by extending the classical percolation models, which can be used to estimate the ESs deployment CAPEX for successful EaaS service. In the derivation of approximations, we obtain an analytical value of critical density for continuum percolation in networks based on the 2D Gilbert disk model, where the maximum communication range between devices is 1\,m, which is in agreement with existing work on percolation theory. %Future extensions of this work include investigating better approximations and discussing regular spatial distributions of ESs.
\appendices
\section{Proof of Lemma~\ref{lem:lowerbound}}\label{app:lowerbound}
When we analyze the sub-critical regime, we assume that $r_f\geq r_r$ and $r_r=a_L$. We consider two special cases: i) there is no IoE device in $\ncalH_L$ or ii) there is at least one IoE device and no ESs inside the outer envelope $\ncalE_{out}^L$. The sufficient condition for no percolation can be expressed as:
\begin{equation}
    % P(\ncalH\ is\ inactive)&>P(No\ device\ in\ \ncalH)+P(At\ least\ one\ device\ in\ \ncalH)P(No\ ES\ in\ \ncalE_{out})\\
\begin{array}{@{}r@{}l}
    \P&\{\ncalH_L \rm{\ is\ inactive}\}\\
    &>e^{-\frac{3\sqrt{3}}{2}\lambda_r a_L^2}+(1-e^{-\frac{3\sqrt{3}}{2}\lambda_r a_L^2})e^{-\lambda_f S_{\ncalE_{out}^L}(r_r,r_f)}>\frac{1}{2}.
\end{array}
\end{equation}
\indent The density of ESs should satisfy:
\begin{equation}
\begin{array}{@{}r@{}l}
    \lambda_f&<\displaystyle\frac{1}{S_{\ncalE_{out}^L}(r_r,r_f)} \ln \frac{1-\exp(-\frac{3\sqrt{3}}{2}\lambda_r a_L^2)}{\frac{1}{2}-\exp(-\frac{3\sqrt{3}}{2}\lambda_r a_L^2)}\\
    &=\displaystyle\frac{1}{S_{\ncalE_{out}^L}(r_r,r_f)} \ln \frac{1-\exp(-\frac{3\sqrt{3}}{2}\lambda_r r_r^2)}{\frac{1}{2}-\exp(-\frac{3\sqrt{3}}{2}\lambda_r r_r^2)},
\end{array}
\end{equation}
where 
\begin{equation}
\displaystyle\frac{1-\exp(-\frac{3\sqrt{3}}{2}\lambda_r r_r^2)}{\frac{1}{2}-\exp(-\frac{3\sqrt{3}}{2}\lambda_r r_r^2)}>0, \,
         \lambda_r>0.
\end{equation}
% \begin{equation}
%     \begin{array}{c}
%     \left\{
%     \begin{array}{c}
%          \displaystyle\frac{1-\exp(-\frac{3\sqrt{3}}{2}\lambda_r r_r^2)}{\frac{1}{2}-\exp(-\frac{3\sqrt{3}}{2}\lambda_r r_r^2)}>0  \\
%          \lambda_r>0 
%     \end{array}
%     \right.
%     \end{array}.
% \end{equation}
% \begin{equation}
%     \lambda_r>\frac{2\ln 2}{3\sqrt{3}r_r^2}\approx \frac{0.267}{r_r^2}.
% \end{equation}
\indent Then, the lower bound of critical ES density can be written as:
\begin{equation}
    \lambda_f^L(\lambda_r,r_r,r_f)=\frac{1}{S_{\ncalE_{out}^L}(r_r,r_f)} \ln \frac{1-\exp(-\frac{3\sqrt{3}}{2}\lambda_r r_r^2)}{\frac{1}{2}-\exp(-\frac{3\sqrt{3}}{2}\lambda_r r_r^2)},
\end{equation}
where
\begin{equation}
    \lambda_r>\frac{2\ln 2}{3\sqrt{3}r_r^2}\approx\frac{0.267}{r_r^2}.
\end{equation}
\indent As shown in Fig.\ref{fig:outerenvelope}, the area of the outer envelope $\ncalE_{out}^L$ can be calculated as:
\begin{equation}
\begin{array}{@{}r@{}l}
    S_{\ncalE_{out}^L}(r_r,r_f)&=\displaystyle 6(\frac{\sqrt{3}}{4}r_r^2+r_r r_f +\frac{\pi}{6}r_f^2)\\
    &=\displaystyle\frac{3\sqrt{3}}{2}r_r^2+6r_r r_f+\pi r_f^2>\pi r_f^2.\\
    %&\in\Big(\pi(r_f+\frac{\sqrt{3}}{2}r_r)^2,\pi(r_f+r_r)^2\Big).
\end{array}
\end{equation}
\indent In the sub-critical regime, we underestimate the probability of an inactive face and underestimate the value of the lower bound. But is still enough for the proof of phase transition. It is worth noting that the lower bound has a minimum value. When $\lambda_r$ approaches $\infty$, the value of $\ln \frac{1-\exp(-\frac{3\sqrt{3}}{2}\lambda_r r_r^2)}{\frac{1}{2}-\exp(-\frac{3\sqrt{3}}{2}\lambda_r r_r^2)}$ approaches $\ln 2$, and $\lambda_f^L$ approaches $\frac{\ln 2}{S_{\ncalE_{out}^L}(r_r,r_f)}$. 
\section{Proof of Lemma~\ref{lem:upperbound}}\label{app:upperbound}
When we analyze the super-critical regime, we assume that $r_f\geq r_r$ and $r_r=\sqrt{13}a_U$. We consider a special case: there is at least one IoE device in $\ncalH_U$ and at least one ES inside the inner envelope $\ncalE_{in}^U$. The sufficient condition for non-zero percolation is expressed as: 
\begin{equation}
\begin{array}{@{}r@{}l}
    \displaystyle \P&\{\ncalH_U \rm{\ is\ active}\}\\
    &>(1-e^{-\frac{3\sqrt{3}}{2}\lambda_r a_U^2})(1-e^{-\lambda_f S_{\ncalE_{in}^U}(r_r,r_f)})>\frac{1}{2}.
\end{array}
\end{equation}
\indent The density of ESs should satisfy:
\begin{equation}
\begin{array}{@{}r@{}l}
    \lambda_f&>\displaystyle-\frac{1}{S_{\ncalE_{in}^U}(r_r,r_f)}\ln \bigg(1-\frac{1}{2-2\exp(-\frac{3\sqrt{3}}{2}\lambda_r a_U^2)}\bigg)\\
    % &=\displaystyle-\frac{1}{S_{\ncalE_{in}^U}}\ln \frac{1-2\exp(-\frac{3\sqrt{3}}{2}\lambda_r a_U^2)}{2-2\exp(-\frac{3\sqrt{3}}{2}\lambda_r a_U^2)}\\
    &=\displaystyle\frac{1}{S_{\ncalE_{in}^U}(r_r,r_f)}\ln \frac{1-\exp(-\frac{3\sqrt{3}}{2}\lambda_r a_U^2)}{\frac{1}{2}-\exp(-\frac{3\sqrt{3}}{2}\lambda_r a_U^2)}\\
    &=\displaystyle\frac{1}{S_{\ncalE_{in}^U}(r_r,r_f)}\ln \frac{1-\exp(-\frac{3\sqrt{3}}{26}\lambda_r r_r^2)}{\frac{1}{2}-\exp(-\frac{3\sqrt{3}}{26}\lambda_r r_r^2)},\\
\end{array}
\end{equation}
where 
\begin{equation}
         \displaystyle\frac{1-\exp(-\frac{3\sqrt{3}}{26}\lambda_r r_r^2)}{\frac{1}{2}-\exp(-\frac{3\sqrt{3}}{26}\lambda_r r_r^2)}>0,\,
         \lambda_r>0.
\end{equation}
% \begin{equation}
%     \begin{array}{c}
%     \left\{
%     \begin{array}{c}
%          \displaystyle\frac{1-\exp(-\frac{3\sqrt{3}}{26}\lambda_r r_r^2)}{\frac{1}{2}-\exp(-\frac{3\sqrt{3}}{26}\lambda_r r_r^2)}>0  \\
%          \lambda_r>0 
%     \end{array}
%     \right.
%     \end{array}.
% \end{equation}
% \begin{equation}
%     \lambda_r>\frac{26\ln 2}{3\sqrt{3}r_r^2}\approx \frac{3.46}{r_r^2}.
% \end{equation}
\indent Then the upper bound of critical ES density can be written as:
\begin{equation}
    \lambda_f^U(\lambda_r,r_r,r_f)=\frac{1}{S_{\ncalE_{in}^U}(r_r,r_f)}\ln \frac{1-\exp(-\frac{3\sqrt{3}}{26}\lambda_r r_r^2)}{\frac{1}{2}-\exp(-\frac{3\sqrt{3}}{26}\lambda_r r_r^2)}, 
\end{equation}
where
\begin{equation}
\lambda_r>\frac{26\ln 2}{3\sqrt{3}r_r^2}\approx \frac{3.46}{r_r^2}.
\end{equation}
\indent As shown in Fig.2, the side length of the hexagon is $a_U$. The inner envelope is composed of six identical arcs with radius $r_f$. Using the law of cosines,
\begin{equation}
    \cos\frac{5\pi}{6}=\frac{a_U^2+b^2-r_f^2}{2a_U b}=-\frac{\sqrt{3}}{2}.
\end{equation}
\indent We get
\begin{equation}
    b=\sqrt{r_f^2-\frac{a_U^2}{4}}-\frac{\sqrt{3}}{2}a_U=\sqrt{r_f^2-\frac{1}{52}r_r^2}-\frac{\sqrt{39}}{26}r_r.
\end{equation}
\indent Using the law of sines, 
\begin{equation}
    \frac{\sin\frac{5\pi}{6}}{r_f}=\frac{\sin\alpha}{b}.
\end{equation}
\indent Next, we have
\begin{equation}
    \alpha=\arcsin\frac{b}{2r_f}.
\end{equation}
\indent The area of the inner envelope is
\begin{equation}
\begin{array}{@{}r@{}l}
S_{\ncalE_{in}^U}(r_r,r_f)&=\displaystyle 6(\frac{\sqrt{3}}{4}b^2+\alpha r_f^2-\frac{1}{2}r_f^2\sin 2\alpha)\\
&=\displaystyle\frac{3\sqrt{3}}{2}b^2+6\alpha r_f^2-3r_f^2\sin 2\alpha<\pi r_f^2.\\
%&\in\Big(\pi(r_f-\frac{1}{\sqrt{13}}r_r)^2,\pi b^2\Big),
\end{array}
\end{equation}
\indent In the super-critical regime, we still underestimate the probability of an active face but overestimate the value of the upper bound. Similar to the lower bound, the upper bound can be also expressed as a non-increasing function of $\lambda_r$, which has a minimum value. When $\lambda_r$ approaches $\infty$, the value of $\ln \frac{1-\exp(-\frac{3\sqrt{3}}{26}\lambda_r r_r^2)}{\frac{1}{2}-\exp(-\frac{3\sqrt{3}}{26}\lambda_r r_r^2)}$ approaches $\ln 2$, and $\lambda_f^U$ approaches $\frac{\ln 2}{S_{\ncalE_{in}^U}(r_r,r_f)}$.

\section{Proof of Lemma~\ref{lem:approximationbound}}\label{app:approximationbound}
\indent In this part, we approximate the critical ES density by imitating the bounds. We compare the sub-critical regime and the super-critical regime in Table.\ref{tab:comparison}. These two bounds are both expressed as: $\frac{1}{S(r_r,r_f)}\ln \frac{1-\exp(-A\lambda_r r_r^2)}{\frac{1}{2}-\exp(-A\lambda_r r_r^2)}$. The main factors are the value of $A$ and the expression of $S(r_r,r_f)$. When $\lambda_r$ is small, the values of bounds are decreasing sharply as $\lambda_r$ increases. When $\lambda_r$ approaches $\infty$, the values of bounds are both expressed as $\frac{\ln 2}{S(r_r,r_f)}$. The trend of critical ES density function $\lambda_f^c(\lambda_r,r_r,r_f)$ is similar to the bounds and the approximations we have obtained, so we propose to assume that the expression of critical density can be written using a similar way. \\
\indent In Lemma \ref{lem:lowerbound} and \ref{lem:upperbound}, we know the limit values of the upper bound and lower bound are expressed using the areas of the outer envelope $\ncalE_{out}^L$ and inner envelope $\ncalE_{in}^U$, respectively. Both these two areas are only relevant to $r_r$ and $r_f$. We define the approximation is $\lambda_f^*(\lambda_r,r_r,r_f)=\frac{1}{S(r_r,r_f)}\ln \frac{1-\exp(-A\lambda_r r_r^2)}{\frac{1}{2}-\exp(-A\lambda_r r_r^2)}$. Considering approximation $\lambda_f^{GD}$, $S(r_r,r_f)$ should satisfy:
\begin{equation}
    \frac{\ln 2}{S(r_r,r_f)}=\frac{\lambda_c(1)}{(2r_f+r_r)^2}.
\end{equation}
\indent Next,
\begin{equation}
    S(r_r,r_f)=\frac{\ln 2}{\lambda_c(1)}(2r_f+r_r)^2.
\end{equation}

 When $\lambda_r$ approaches $\frac{\ln 2}{A r_r^2}$, the value of $\lambda_f^*(\lambda_r,r_r,r_f)$ approaches $\infty$. To find the suitable value of $A$, we adopt the necessary condition for percolation: $\lambda_r>\frac{\lambda_c(1)}{r_r^2}$. Let
\begin{equation}
    \frac{\ln 2}{A r_r^2}=\frac{\lambda_c(1)}{r_r^2}.
\end{equation}
\indent Next, we have
\begin{equation}
    A=\frac{\ln 2}{\lambda_c(1)},
\end{equation}
where $0.768<\lambda_c(1)<3.37$. Considering the sub-critical regime and super-critical regime, $A=\frac{\ln 2}{\lambda_c(1)}\in (\frac{3\sqrt{3}}{26},\frac{3\sqrt{3}}{2})$ and $\lambda_c(1)\in(0.267,3.46)$. In some simulation-based approximations in literature \cite{quintanilla2000efficient}, $\lambda_c(1)\approx 1.44$. But in this part, we obtain the theoretical value of $\lambda_c(1)$ for percolation in the infinite wireless network without simulation but just using theoretical analysis.\\
\indent The expressions of areas of $\ncalE_{out}^L$ and $\ncalE_{in}^U$ are shown in (\ref{souterL}) and (\ref{sinnerU}), which are also compared in Table.\ref{tab:comparison}. When $r_f\gg r_r$, $S_{\ncalE_{out}^L}\sim\pi r_f^2$ and $S_{\ncalE_{in}^U}\sim\pi r_f^2$. When we set $\gamma=\frac{r_r}{r_f}$, the areas can be expressed as a function of $r_f$ and $\gamma$.  They can be both considered as the product of $\pi r_f^2$ and a function of $\gamma$.\\
\indent For the sub-critical regime:
\begin{equation}
    S_{\ncalE_{out}^L}(r_r,r_f)=r_f^2(\frac{3\sqrt{3}}{2}\gamma^2+6\gamma+\pi)>\pi r_f^2.
\end{equation}
\indent For the super-critical regime:
\begin{equation}
    S_{\ncalE_{in}^U}(r_r,r_f)=r_f^2(\frac{3\sqrt{3}}{2}c^2+6\alpha-3\sin 2\alpha)<\pi r_f^2,
\end{equation}
where $c=\frac{b}{r_f}=\sqrt{1-\frac{1}{52}\gamma^2}-\frac{\sqrt{39}}{26}\gamma$ and $\alpha=\arcsin \frac{c}{2}$.

\indent We define:
\begin{equation}
    g_L(\gamma)=\frac{S_{\ncalE_{out}^L}(r_r,r_f)}{\pi r_f^2},
\end{equation}
and 
\begin{equation}
    g_U(\gamma)=\frac{S_{\ncalE_{in}^U}(r_r,r_f)}{\pi r_f^2}.
\end{equation}
\begin{figure}
    \centering
    \includegraphics[width=1\columnwidth]{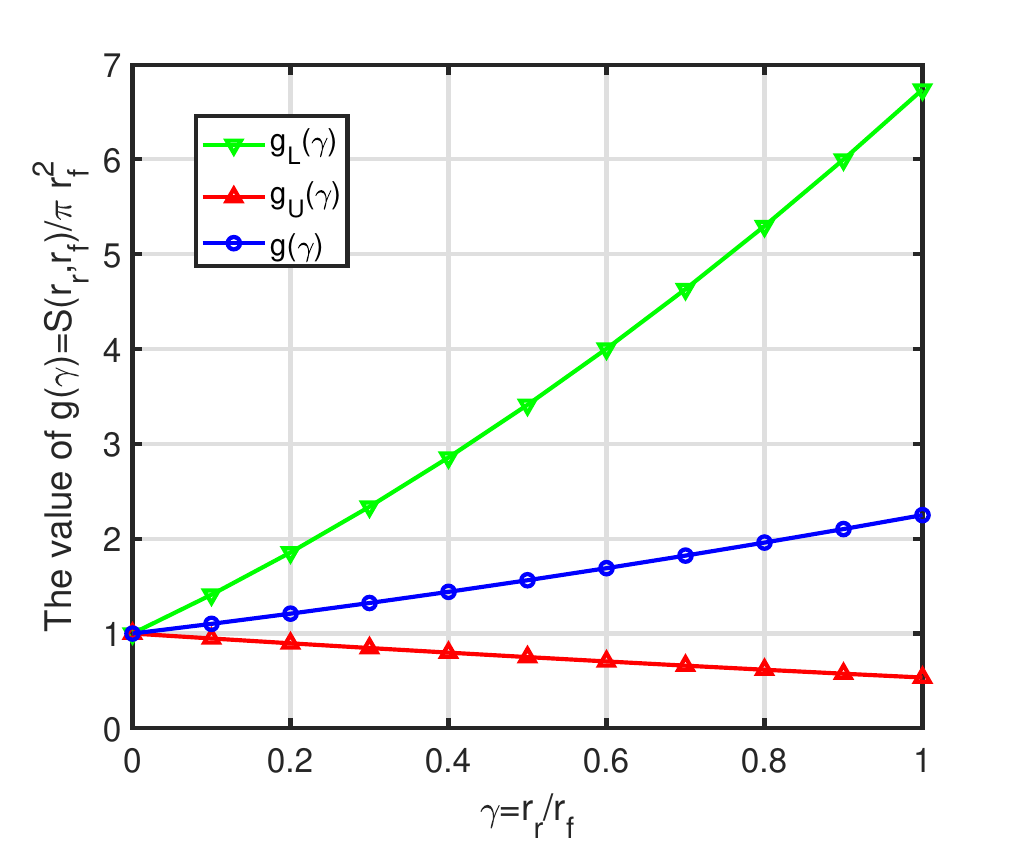}
    \caption{Comparison between $g_U(\gamma)$, $g_L(\gamma)$ and $g(\gamma)$.}
    \label{fig:g_gamma}
\end{figure}

\indent As shown in Fig.\ref{fig:g_gamma}, $g_L(\gamma)$ and $g_U(\gamma)$ both approach 1 when $r_r\ll r_f(\gamma=0)$.  We assume that 
\begin{equation}
    g(\gamma)=\frac{S(r_r,r_f)}{\pi r_f^2} .
\end{equation}
\indent Due to $\lambda_f^L<\lambda_f^c<\lambda_f^U$, $S_{\ncalE_{in}^U}(r_r,r_f)<S(r_r,r_f)<S_{\ncalE_{out}^L}(r_r,r_f)$ and $g_U(\gamma)<g(\gamma)<g_L(\gamma)$. Because $g_L(0)=g_U(0)=1$, %At the same time,   the lower bound and the upper bound both have the same values when $\lambda_r\uparrow \infty$.  We haven't found the exact expression of $g(\gamma)$. For convenience, we let $g(\gamma)=(1+k\gamma)^2$, where $g(0)=1$. \\
% \indent To get the value of $k$, we consider the special case introduced in Lemma \ref{lem:firstapproximation}. The limit value of $\lambda_f^*$ should not exceed $\lambda_f^2$, \ie
% \begin{equation}
%     \frac{\ln 2}{\pi r_f^2 (1+k \gamma)^2}\leq \frac{\lambda_c(1)}{(2r_f+r_r)^2}.
% \end{equation}
% \indent Then
% \begin{equation}
%     k\geq \max \left\{\bigg(\sqrt{\frac{4\ln 2}{\pi \lambda_c(1)}}-1\bigg) \frac{1}{\gamma}+\sqrt{\frac{\ln 2}{\pi \lambda_c(1)}}\right\}=\sqrt{\frac{9\ln 2}{\pi \lambda_c(1)}}-1\approx 0.174.
% \end{equation}
\begin{equation}
\begin{array}{@{}r@{}l}
    &g(\gamma)|_{\gamma=0}=\displaystyle\frac{S(r_r,r_f)}{\pi r_f^2}|_{r_f\gg r_r}\\
    &=\displaystyle\frac{\ln 2}{\pi\lambda_c(1)}\frac{(2r_f+r_r)^2}{r_f^2}|_{r_f\gg r_r}\\
    &=\displaystyle\frac{\ln 2}{\pi\lambda_c(1)}(2+\gamma)^2|_{\gamma=0}=\displaystyle\frac{4\ln 2}{\pi\lambda_c(1)}=1.
\end{array}
\end{equation}
\indent Then, we have
\begin{equation}
    \lambda_c(1)=\frac{4\ln 2}{\pi}\approx 0.883<1.44.
\end{equation}

\indent To prove the feasibility of $\lambda_f^*$, we compare $\lambda_f^*$ with $\lambda_f^{IC}$: 
\begin{equation}
    \frac{\lambda_f^*(\lambda_r,r_r,r_f)}{\lambda_f^{IC}(\lambda_r,r_r,r_f)}=\frac{\frac{1}{\displaystyle S(r_r,r_f)}\ln\displaystyle \frac{1-\exp(-\frac{\ln 2}{\lambda_c(1)}\lambda_r r_r^2)}{\frac{1}{2}-\exp(-\frac{\ln 2}{\lambda_c(1)}\lambda_r r_r^2)}}{-\displaystyle\frac{1}{\pi r_f^2} \ln (1-\frac{\lambda_c(1)}{r_r^2})}.
\label{approandappro1}
\end{equation}
\indent In (\ref{approandappro1}),
\begin{equation}
    \lim_{\lambda_r\rightarrow (\frac{\lambda_c(1)}{r_r^2})^{+}}\frac{\ln \displaystyle\frac{1-\exp(-\frac{\ln 2}{\lambda_c(1)}\lambda_r r_r^2)}{\frac{1}{2}-\exp(-\frac{\ln 2}{\lambda_c(1)}\lambda_r r_r^2)}}{-\ln (1-\frac{\lambda_c(1)}{r_r^2})}=1.
\end{equation}
\indent Then,
\begin{equation}
    \frac{\lambda_f^*(\lambda_r,r_r,r_f)}{\lambda_f^{IC}(\lambda_r,r_r,r_f)}\bigg|_{r_f\gg r_r}=\frac{\pi r_f^2}{S(r_r,r_f)}\bigg|_{r_f\gg r_r}=1,
\end{equation}
% and 
% \begin{equation}
%     \frac{\frac{1}{\pi r_f^2}(1+k\gamma)^2}{\frac{1}{\pi r_f^2}}
% \end{equation}
which means when $r_f$ is much larger than $r_r$, the $\lambda_f^*$ and $\lambda_f^{IC}$ are totally matching. 

\ifCLASSOPTIONcaptionsoff
  \newpage
\fi

\bibliographystyle{IEEEtran}
\bibliography{ref}
\end{document}